\documentclass[a4paper,11pt]{article}
\pdfoutput=1 

\usepackage{jheppub}

\usepackage{amsmath,graphicx,amssymb,subfigure,bbm}
\usepackage[all]{xy}
\preprint{DIAS--STP--14--xx}

\title{\boldmath Holographic Bilayer/Monolayer Phase Transitions}

\author[a]{Veselin G. Filev,}
\author[b]{Matthias Ihl,}
\author[b]{and Dimitrios Zoakos}
\affiliation[a]{School of Theoretical Physics,\\ 
       Dublin Institute for Advanced Studies, \\
       10 Burlington Road, 
       Dublin 4, Ireland.}
\affiliation[b]{Centro de F\'isica do Porto \&  Departamento de F\'isica e Astronomia\\ 
Faculdade de Ci{\^e}ncias da Universidade do Porto,\\ Rua do Campo Alegre 687, 4169-007 Porto, Portugal.}       
\emailAdd{vfilev@stp.dias.ie}
\emailAdd{matthias.ihl@fc.up.pt}
\emailAdd{dimitrios.zoakos@fc.up.pt}

\abstract{In the present work, we discuss the phase structure of bilayer and monolayer phases in the (2+1)-dimensional defect field theory whose gravity dual is
obtained by embedding  $D5$/anti-$D5$ probe flavour branes in the singular conifold. We study in detail the embedding equations and compare the free energies of 
the resulting configurations at non-zero temperature and external magnetic field perpendicular to the defects. 
Moreover, we analyse the meson spectrum and confirm the stability of the single bilayer solution.}

\begin{document}
\maketitle
\flushbottom


\section{Introduction}
\label{sec:intro}
It has been realised in the last few years that holographic brane constructions can be successfully used to study condensed matter systems in the strong coupling regime. Many features of a growing number of models bear a surprisingly close resemblance to actual physical systems. Examples include high-$T_c$ superconductivity, strange metals, and quantum liquids of the marginal and non-Fermi type. The probe brane configuration employed in this paper leads to a (2+1)-dimensional defect field theory of strongly coupled fermions living on domain walls in the (3+1)-dimensional ambient space-time, and is potentially useful to describe the physics of monolayers (one sheet) and bilayers (two sheets) of graphene. The specific construction under consideration, which is a generalisation of the situation studied in \cite{Filev:2013vka}, involves $N_f \ll N_c$ $D5/\overline{D5}$-probe brane pairs in the Klebanov-Witten background generated by $N_c$ $D3$ branes located at the tip of the 
singular conifold \cite{Klebanov:1998hh}. In \cite{Filev:2013vka}, the focus was on studying an $r$-dependent profile of the $D5/\overline{D5}$-brane embedding in the internal space (thus corresponding to a monolayer phase) which geometrically realises spontaneous conformal \cite{Ben-Ami:2013lca} and chiral symmetry breaking via a U-shaped brane embedding similar to the Kuperstein-Sonnenschein $D3/D7/\overline{D7}$ construction \cite{Kuperstein:2008cq}. There is also the possibility of straight embeddings that fall into the horizon of the black hole once a non-zero temperature is turned on. An interesting competition between U-shaped and straight embeddings corresponding to chiral/conformal symmetry broken and restored phases, resp., can be observed by turning on world volume gauge fields on the probe brane corresponding to external (electro-)magnetic fields \cite{Filev:2013vka, Alam:2012fw}. In the present paper, we generalise the aforementioned model by considering the possibility of an $r$-dependent 
profile $z(r)$ in the $x^3$-direction transverse to the (domain wall) defects, at finite temperature and magnetic field.\footnote{See \cite{Evans:2013jma} for a recent, related (2+1)-dimensional bilayer construction in the $AdS_5 \times S^5$ background.} In this way, we obtain an even richer phase structure of single and combined bilayer and monolayer phases of our (2+1)-dimensional model. Namely, we study the embedding equations obtained from the DBI action of the $D5/\overline{D5}$-branes, and compare the free energies of the resulting configurations. This construction can potentially play a r\^ole in improving the understanding of dynamical symmetry breaking in graphene monolayers and bilayers, see e.g., \cite{Gorbachev:2012} and references therein; the reader should also consult \cite{Grignani:2012qz} for a recent related holographic model involving $D7$-branes, and a more detailed explanation of the relevance of holographic setups to learn some lessons about graphene bilayers.

The paper is organised as follows: In section \ref{sec:setup}, we present the general setup of the brane construction, mainly repeating the essentials from  \cite{Filev:2013vka}. 
Then, in section \ref{sec:singlebilayer}, we study in detail the single bilayer phase at zero and non-zero temperature as well as an external magnetic field and present the 
bilayer/monolayer phase transition. 
This is followed in section \ref{sec:spectrum} by a thorough investigation of the corresponding meson spectrum and its stability analysis. 
Finally, in section \ref{sec:combined}, we study the combined bilayer and monolayer phase at zero and non-zero temperature and finite magnetic field.


\section{General setup}
\label{sec:setup}

Let us consider the Klebanov-Witten background, i.e., type IIB supergravity on $AdS_5 \times T^{1,1}$ space-time, as the near-brane
geometry generated by $N_c$ $D3$-branes placed at the tip of the conifold singularity \cite{Klebanov:1998hh}. The metric is given by
\begin{align}\label{eq:metric}
d s^2 &= \frac{r^2}{L^2}\left(-dt^2 + d x_1^2 + d x_2^2 +d x_3^2\right) \nonumber \\
&\quad + \frac{L^2}{r^2}\left[{dr^2} + \frac{r^2}{6} \left( \sum_{i=1}^2 d \theta_i^2 + {\rm sin}^2 \theta_i d \phi_i^2 \right)+ \frac{r^2}{9}\left(d \psi + \sum_{i=1}^2 {\rm cos} \theta_i d \phi_i \right)^2\right], 
\end{align}
where $L^4=\frac{27}{4}\pi g_s N_c l_s^4$ and the range of angles is $0<\theta_{1,2}<\pi$, $0<\phi_{1,2}<2 \pi$ and $0<\psi< 4 \pi$.
Moreover, we will introduce $N_f \ll N_c$ flavour probe $D5/\overline{D5}$-brane pairs such that (2+1)- dimensional fundamental degrees of freedom are added to the quiver diagram of the theory. This means that 
the flavour and the colour branes have to intersect in a (2+1)-dimensional defect in the (3+1)-dimensional ambient Minkowski space-time. The case where the flavour branes are located at $x_3= \mathrm{const.}$ was recently constructed and investigated  in \cite{Filev:2013vka, Ben-Ami:2013lca}.\\
Here, we will study a generalisation of this construction, when $x_3$ is also allowed to describe a profile in $r$, which leads to the following ansatz for the U-shaped embeddings:
\begin{equation*}
  \begin{array}{|c||c|c|c|c|c|c|c|c|c|c|}
    \hline
     & x^0 & x^1 & x^2 & x^3 & r & \theta_- & \phi_+ & \theta_+ & \phi_- & \psi \\
    \hline
    {\rm D3} & \times & \times & \times & \times & \cdot & \cdot & \cdot & \cdot & \cdot & \cdot \\
    \hline
    {\rm D5}/\overline{\rm D5} & \times & \times & \times  & z(r) & \times & \times & \times &\cdot & \cdot & \psi(r) \\ \hline
  \end{array}
\end{equation*}
Using the notations of \cite{Filev:2013vka}, we have defined
\begin{equation}
\theta_\pm \, = \, \frac{\theta_1 \pm \theta_2}{2} \quad \mathrm{and} \quad \phi_\pm \, = \, \frac{\phi_1 \pm \phi_2}{2},
\end{equation}
and we henceforth fix, without loss of generality, $\theta_{-}=0\, , \,  \phi_{+}=\pi$.


\section{Single bilayer}
\label{sec:singlebilayer} 

\begin{figure}[h] 
   \centering
   \includegraphics[width=4in]{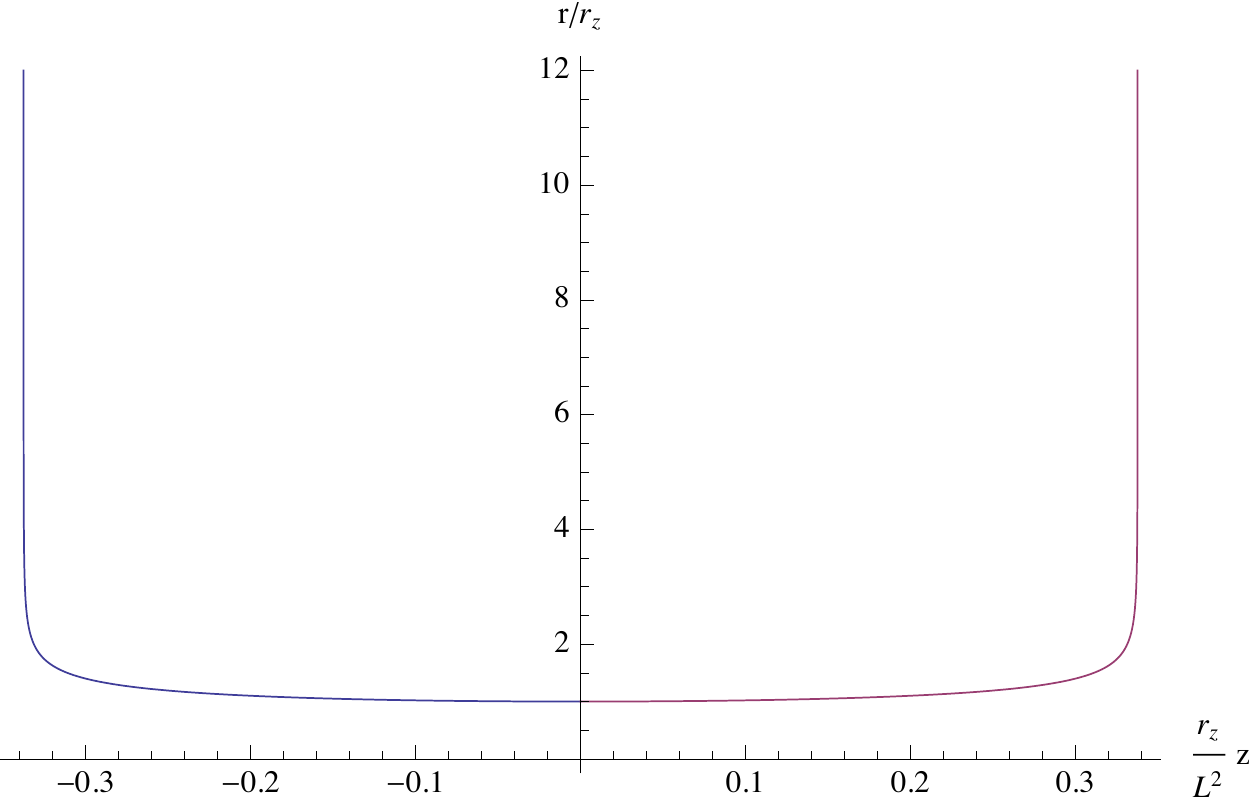} 
   \caption{Analytic $D5/\overline{D5}$-brane U-shaped embedding with profile in the $z$ and $r$.}
   \label{fig:singel-1}
\end{figure}

We will start our discussion by studying in detail the single bilayer configuration, for which $x^3=x^3(r)=:z(r)$ describes a profile in $r$ and $\psi = \mathrm{const}.$, at finite temperature and external magnetic field perpendicular to the $(2+1)$-dimensional defect. With this ansatz the induced metric on the world volume of the D5--brane configuration is:
 \begin{align}\label{eq:D5metric-single}
ds^2=\frac{r^2}{L^2}\left(-dt^2+dx_1^2+dx_2^2\right)+\frac{L^2}{r^2}\left[dr^2\left(1+\frac{r^4}{L^4}z'(r)^2\right)+\frac{r^2}{3}d\Omega_2^2 \right]
\end{align}
and the corresponding DBI action becomes:
\begin{equation}\label{eq:DBI-single}
S_{\mathrm{D5}}=- \tau_5 \int d \xi^6 \sqrt{\det P[g]}=- 2 \mathcal{N} \int dr \, r^2 \sqrt{1+\frac{r^4}{L^4}z'(r)^2}\ ,
\end{equation}
where $\mathcal{N}= \frac{2 \pi}{3}\tau_5 \mathrm{Vol}(\mathbb{R}^{2,1})$. Integrating once the equation of motion, for $z$ we obtain
\begin{align}\label{1st-integ-z}
\frac{\frac{r^6}{L^4}z'(r)}{\sqrt{1+\frac{r^4}{L^4}z'(r)^2}}&= \, \frac{r_z^4}{L^2}\ ,
\end{align}
where $r_z$ is the minimum value of $r$ that the U-shaped embedding reaches, i.e., the position where the two branches ($D5$ and $\overline{D5}$ brane) merge.
The resulting profile for $z(r)$ is
\begin{equation}\label{profile-z}
z(r)=\pm\frac{L^2}{r_z}\left(\frac{\pi^{1/2}\Gamma(\frac{5}{8})}{\Gamma(\frac{1}{8})}+\frac{r_z^5}{5r^5}\,{}_2F_1\left[\frac{1}{2}\ ,\frac{5}{8}\ ,\frac{13}{8}\ ,\frac{r_z^8}{r^8}\right]\right)\ ,
\end{equation}
where the two choices of the overall sign correspond to the two branches of the U-shaped embedding (look at figure \ref {fig:singel-1}).
For the large $r$ expansion of $z$ we obtain:
\begin{equation}\label{expansionz}
z(r)=\pm\left(\frac{\pi^{1/2}\Gamma(\frac{5}{8})}{\Gamma(\frac{1}{8})}\,\frac{L^2}{r_z}-\frac{r_z^4L^2}{5r^5}+O(r^{-13})\right)\ .
\end{equation}
This implies that in the limit $r\to\infty$ the branches of the U-shaped embedding approach two straight brane D5-brane embeddings corresponding to a set of defect field theories localised at $z=-\frac{\pi^{1/2}\Gamma(\frac{5}{8})}{\Gamma(\frac{1}{8})}\,\frac{L^2}{r_z}$ and 
$z=\frac{\pi^{1/2}\Gamma(\frac{5}{8})}{\Gamma(\frac{1}{8})}\,\frac{L^2}{r_z}$, respectively. Thus this is a bi-layer configuration with separation between the layers $\Delta z$ given by:
\begin{equation}\label{Deltaz}
\Delta z =\frac{2\pi^{1/2}\Gamma(\frac{5}{8})}{\Gamma(\frac{1}{8})}\,\frac{L^2}{r_z}.
\end{equation}
Furthermore, if we denote schematically the fundamental fields on the two defects as $\psi_+$ and $\psi_-$, correspondingly, the AdS/CFT dictionary relates the bi-layer condensate $\langle {\cal O}_{\psi_+\psi_-}\rangle$ sourced by $\Delta z$ to the radial distance $r_z$ via:
\begin{equation}\label{condz}
\langle{\cal O}_{\psi_+\psi_-}\rangle \propto c_z = -\frac{r_z^4L^2}{5} \ .
\end{equation}
The condensate $\langle{\cal O}_{\psi_+\psi_-}\rangle$ breaks conformal symmetry and also breaks some of the global  $U(N_f)\times U(N_f)$ flavour symmetry of the theory, since the operator ${\cal O}_{\psi_+\psi_-}$ couples the fields $\psi_+$ and $\psi_-$. While we do not know the explicit form of the operator ${\cal O}_{\psi_+\psi_-}$, we are still able to determine its conformal dimension. According to the standard AdS/CFT dictionary, the conformal dimension is encoded in the exponent of the corresponding supergravity field $z(r)$ as $r\to\infty$. For a three dimensional field theory, the leading mode should behave as  $r^{\Delta-3+p}$, while the subleading mode should asymptote to $r^{-\Delta+p}$ for some constant $p$. Using equation (\ref{expansionz}), one can easily check that the operator ${\cal O}_{\psi_+\psi_-}$ has conformal dimension four. Using equations (\ref{Deltaz}) and (\ref{condz}), one can verify that $\langle{\cal O}_{\psi_+\psi_-}\rangle \propto 1/(\Delta z)^4$, which agrees with the 
operator ${\cal O}_{\psi_+\psi_-}$ having engineering dimension four. 

Note that if we set $r_z=0$, the analogue of the U-shaped embedding is given by a pair of parallel $D5/ \overline{D5}$-branes positioned at $z=\pm\frac{\pi^{1/2}\Gamma(\frac{5}{8})}{\Gamma(\frac{1}{8})}\,\frac{L^2}{r_z}$ for all $r$. In this case, the two embeddings represent independent domain walls, and there is no condensate coupling the two defect field theories; the theory is in a mono-layer phase. For a given separation between the domain walls, $\Delta z$, this configuration competes with the U-shaped embedding. To determine which phase of the theory is preferred, we have to compare the free energies of each phase. The free energy is proportional to the regularised wick rotated on-shell action of the system. One can show that the on-shell action diverges as $\Lambda_{UV}^3$, where $\Lambda_{UV}$ is a UV cut-off. To regularise the on-shell action, it is sufficient to add a volume counter term \cite{Karch:2005ms} proportional to $\int\sqrt{\gamma}$ at $r=r_{max}\propto \Lambda_{UV}$. Using equations (\ref{1st-integ-z}) and (\ref{eq:DBI-single}) for the free energies of the U-shaped and parallel embeddings we obtain
\begin{eqnarray}
 F_{U}&=&2\,{\cal N'}\left(\,\int\limits_{r_z}^{\infty}dr\left(\,\frac{r^6}{\sqrt{r^8-r_z^8}}-r^2\right)-\frac{r_z^3}{3}\right) =\,\frac{{\cal N'}\,\sqrt{\pi}\, \Gamma\left(-\frac{3}{8}\right) \,  r_z^3}{4 \,\Gamma\left(\frac{1}{8}\right)}\,<\,0 \ ,  \\
 F_{||}&=0&\ ,
\end{eqnarray}
where ${\cal N'}= \frac{2 \pi}{3}\tau_5 \mathrm{Vol}(\mathbb{R}^2)$.
We conclude that $F_{U}<F_{||}$ and therefore the U-shaped embeddings are preferred to the parallel ones. This suggests that, at zero temperature, the theory is always in the bilayer phase.

Note that the only independent scale in the theory is the separation between the two domain walls, $\Delta z$. Since the underlying theory is conformal, it is not a surprise that all physical quantities  can be expressed in terms of the energy scale $1/\Delta z$ associated with this separation. The situation will be different  if we introduce an extra physical scale such as temperature or magnetic field.


\subsection{Bilayer/monolayer thermal phase transition}

In this subsection we investigate the effect of finite temperature on the bilayer configuration studied above. Intuitively, we expect that above some critical temperature $T_c\propto 1/\Delta z$ the bilayer condensate will melt and the theory will be in a mono-layer phase. To turn on a temperature, we substitute the AdS$_5$ part of the geometry with an AdS-black hole. Then the temperature of the dual gauge theory is given by the temperature of the black hole. For the induced metric on the world volume of the $D5$-brane we obtain
 \begin{align}\label{eq:D5metric-Tsingle}
ds^2=\frac{r^2}{L^2}\left(-b(r)\,dt^2+dx_1^2+dx_2^2\right)+\frac{L^2}{r^2\,b(r)}\left[dr^2\left(1+\frac{r^4\,b(r)}{L^4}\,z'(r)^2\right)+\frac{r^2}{3}d\Omega_2^2 \right]\ ,
\end{align}
where the emblackening factor is $b(r)=1-\frac{r_H^4}{r^4}$. \\
The corresponding DBI action and  equation of motion become:
\begin{equation}\label{eq:DBI-single-T}
S_{\mathrm{D5}}=- \tau_5 \int d \xi^6 \sqrt{\det P[g]}=- 2 \mathcal{N} \int dr \, r^2 \sqrt{1+\frac{r^4\,b(r)}{L^4}z'(r)^2}\ ,
\end{equation}
\begin{equation}\label{1st-int-zT}
\frac{\frac{r^6\,b(r)}{L^4}z'(r)}{\sqrt{1+\frac{r^4\,b(r)}{L^4}z'(r)^2}}= \Pi^T_z\ .
\end{equation}
Next defining $r_z$ such that
\begin{equation}
r_z^4\,\sqrt{b(r_z)}\,= \, \Pi^T_z\,L^2\ \,~~~~\text{for}~r_z\geq r_H
\end{equation}
we find
\begin{align}
z(r)&=\pm\int\limits_{r_z}^{r}d\bar{r}\,\frac{r_z^4\,\sqrt{b(r_z)}\,L^2}{\bar{r}^2\,\sqrt{b(\bar{r})}\sqrt{\bar{r}^8\,b(\bar{r})-r_z^8\,b(r_z)}}; &\text{for}~r_z>r_H\ , \label{U-T}\\
z(r)&=\text{const}=\pm\frac{\Delta z}{2};  &\text{for}~r_z=r_H\ , \label{parallel-T}
\end{align}
where $\Delta z= 2|z(\infty)|$. The large $r$ expansion of $z$ yields
\begin{equation}\label{expand-z}
z(r)=\pm\left(\frac{\Delta z}{2} -\frac{r_z^4\,\sqrt{b(r_z)}\,L^2}{5\,r^5}+O(1/r^{9})\right)\ ,
\end{equation}
therefore the bilayer condensate is given by
\begin{equation}\label{condzT}
\langle{\cal O}_{\psi_+\psi_-}\rangle \propto c_z = -\frac{L^2}{5}\,r_z^2\,\sqrt{r_z^4-r_H^4} \ .
\end{equation}
We see that at $r_z=r_H$ the condensate vanishes and the parallel embeddings given by equation (\ref{parallel-T}) correspond to two non-interacting domain walls -- a monolayer phase. To determine the stable phase we have to compare the free energies of the U-shaped (\ref{U-T}) and parallel (\ref{parallel-T}) embeddings. Using the same regularisation as in the zero temperature case, the free energies are
\begin{align}
F_{U}&= 2\,{\cal N'}\left(\,\int\limits_{r_z}^{\infty}dr\left(\,\frac{r^6\,\sqrt{b(r)}}{\sqrt{r^8\,b(r)-r_z^8\,b(r_z)}}-r^2\right)-\frac{r_z^3}{3}\right) ,~~~\text{for}~r_z>r_H\ \label{F-U-T} \\
F_{||}&=2\,{\cal N'}\left(\,\int\limits_{r_H}^{\infty}dr\left(r^2-r^2\right)-\frac{r_H^3}{3}\right)=-\frac{2\,{\cal N'}}{3}\,r_h^3\ . 
\end{align}
To compare the free energies we have to evaluate numerically the free energy of the bilayer phase (\ref{F-U-T}). To this end it is convenient to define the dimensionless variables,
\begin{equation}
\tilde r =\frac{r}{r_H} ;~~\Delta\tilde{z}=\frac{r_H}{L^2}\,\Delta z;~~\tilde r_z=\frac{r_z}{r_H};
\end{equation}
The free energy is then given by $F ={\cal N'}\,r_H^3\,\tilde F(\tilde r_z)$. As expected it scales as $T^3$. Clearly for the parallel embeddings $\tilde F_{||}=-2/3$, while for the U-shaped embeddings we have to calculate $\tilde F_{U}$ numerically as a function of the parameter $\tilde r_z$. Since $\tilde r_z$ is not a bare parameter of the dual gauge theory (it is related to the condensate), we express $\tilde r_z= \tilde r_z(\Delta \tilde z)$ and plot the free energy $\tilde F \propto \frac{F}{T^3}$ as a function of $\Delta \tilde z \propto \Delta z \,T$.

\begin{figure}[h] 
   \centering
   \includegraphics[width=7.5cm]{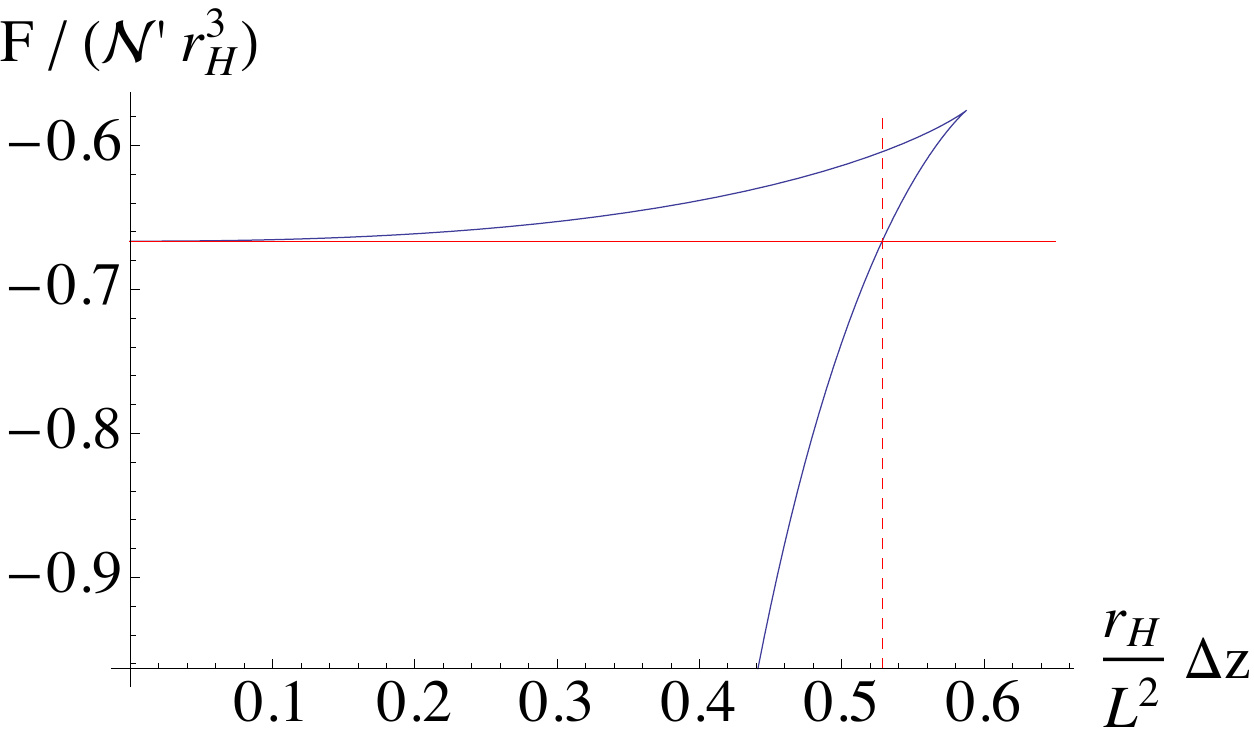} 
   \includegraphics[width=7.5cm]{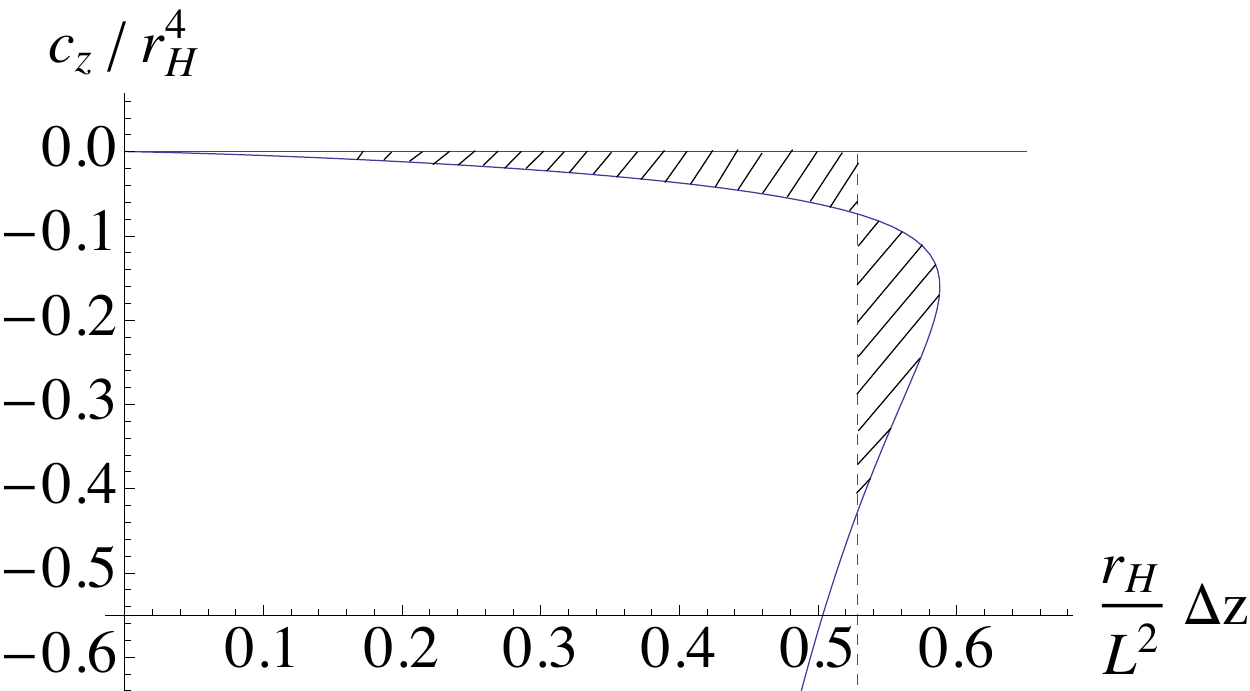} 
   \caption{}
   \label{fig:2-T}
\end{figure}

In figure \ref{fig:2-T}, we present plots of the free energy and the condensate $\tilde c_z$ as functions of  the separation between the layers, $\Delta z$. The blue curve in the first plot from left to right, represents the free energy in the bilayer phase, while the red line represents the free energy in the monolayer phase. The vertical dashed line represents the critical value of $\Delta\tilde z$, for which the free energies are the same. One can see that at ${\Delta\tilde z}_{cr}=r_H\,\Delta z/L^2\approx 0.5285$ there is a first order phase transition from the bilayer phase to the monolayer phase. Since $T=r_H/(\pi L^2)$, we find the following result for the critical temperature,

\begin{equation}\label{Tcr}
T_{cr}=\frac{({\Delta\tilde z}_{cr}/\pi)}{\Delta z} \approx \frac{0.1682}{\Delta z}\ .
\end{equation}
As one expects from conformality, the critical temperature is set by the only other energy scale in the theory, $1/\Delta z$. In the second plot in figure \ref{fig:2-T}, the blue curve represents the bilayer condensate, while the red line represents the vanishing condensate in the monolayer phase. The dashed vertical line again represents the critical value ${\Delta\tilde z}_{cr}$. One can see that at the phase transition there is a finite jump of the condensate. Furthermore, one can verify that the shaded regions in the plot have equal areas, which is consistent with Maxwell's equal area law.  

Note that while, technically the bilayer U-shaped configuration that we considered in this section is an analogue of the U-shaped embedding in the $(r,\psi)$-plane considered in ref.~\cite{Filev:2013vka},  the physics described by these configurations is completely different. Indeed, at finite temperature the parallel embeddings are always preferred relative to the U-shaped embeddings in the $(r,\psi)$-plane. Here, by contrast, we have uncovered a first order phase transition for the bilayer configuration. The intuitive explanation for this difference is that, while in the bilayer case the parameter $\Delta z$ is dimensionful and defines an energy scale $1/\Delta z$, for the configuration considered in ref.~\cite{Filev:2013vka},  the analogous parameter $\Delta\psi$ is dimensionless and there is no energy scale associated to it that could set the critical temperature (as in equation (\ref{Tcr})). The situation changes if one introduces an extra scale into the theory, such as magnetic field~\cite{Filev:2013vka}, which breaks conformality even at zero temperature. In the next subsection we consider the single bilayer in the presence of both finite temperature and magnetic field. 

\subsection{Single bilayer at finite temperature and magnetic field}\label{SBLTB}

In order to excite an external magnetic field, we turn on a $U(1)$ gauge field on the probe branes. To this end, we consider the ansatz $A_2 = \frac{B}{2\pi\alpha'} x_1$, which corresponds to a constant magnetic field $F_{12} =  {B}/{(2\pi\alpha')}$ along the $x_3$ direction, perpendicular to the defect. For the DBI action and the first integral of the equation of motion we obtain
\begin{equation}\label{eq:DBI-TBsingle}
S_{\mathrm{D5}}=- 2 \mathcal{N}_T \int dr \, \sqrt{r^4+B^2L^4}\, \sqrt{1+\frac{r^4\,b(r)}{L^4}\,z'(r)^2}\ ,
\end{equation}
where  $ \mathcal{N}_T = \frac{{\cal N'}}{T}= \frac{2 \pi}{3}\tau_5 \mathrm{Vol}(\mathbb{R}^2)/T$, and 
\begin{equation}
\frac{\frac{r^6}{L^4} \sqrt{1 + B^2 \frac{L^4}{r^4}} b(r) z'(r)}{\sqrt{1+\frac{r^4b(r)}{L^4}z'(r)^2}}= \Pi^{T,B}_{z} \ .\\
\end{equation}
Defining $r_B=\sqrt{B}\,L$ and $\Pi^{T,B}_z:= \frac{r_z^4}{L^2}\sqrt{b(r_z)}\sqrt{1+\frac{r_B^4}{r_z^4}}$, we find the following result for the profile of $z$ ,
\begin{align}
z(r)&=\pm\int\limits_{r_z}^{r}d{r}\,\frac{L^2\,r_z^2\,\sqrt{b(r_z)}\,\sqrt{r_z^4+r_b^4}}{r^2\,\sqrt{b({r})}\sqrt{r^4\,({r}^4+r_B^4)\,b({r})-r_z^4\,(r_z^4+r_B^4)\,b(r_z)}}; &\text{for}~r_z>r_H\ , \label{U-TB}\\
z(r)&=\text{const}=\pm\frac{\Delta z}{2};  &\text{for}~r_z=r_H\ , \label{parallel-TB}
\end{align}
where $\Delta z=2 |z(\infty)|$. For the large $r$ expansion of $z$ we obtain
\begin{equation}\label{expand-z}
z(r)=\pm\left(\frac{\Delta z}{2} -\frac{L^2\,\sqrt{r_z^4-r_H^4}\sqrt{r_B^4+r_z^4}}{5\,r^5}+O(1/r^{9})\right)\ .
\end{equation}
Therefore the bilayer condensate is
\begin{equation}\label{condzT}
\langle{\cal O}_{\psi_+\psi_-}\rangle \propto c_z = -\frac{L^2}{5}\,\sqrt{r_z^4-r_H^4}\sqrt{r_B^4+r_z^4} \ .
\end{equation}
Note that the condensate decreases as $r_H$ (the temperature) is increased. At $r_H=r_z$ the bilayer condensate vanishes and the theory is in a monolayer phase containing two parallel domain walls described by parallel D5--brane embeddings. On the other hand, we see that the magnetic field enhances the condensate (its absolute value grows with $r_B$). This is an expected behaviour, due to the universal nature of the effect of magnetic catalysis of chiral symmetry breaking. The competition between the dissociating effect of the temperature and the binding effect of the external magnetic field results in an interesting phase diagram, which we analyse bellow. 

\begin{figure}[h] 
   \centering
   \includegraphics[width=5.0in]{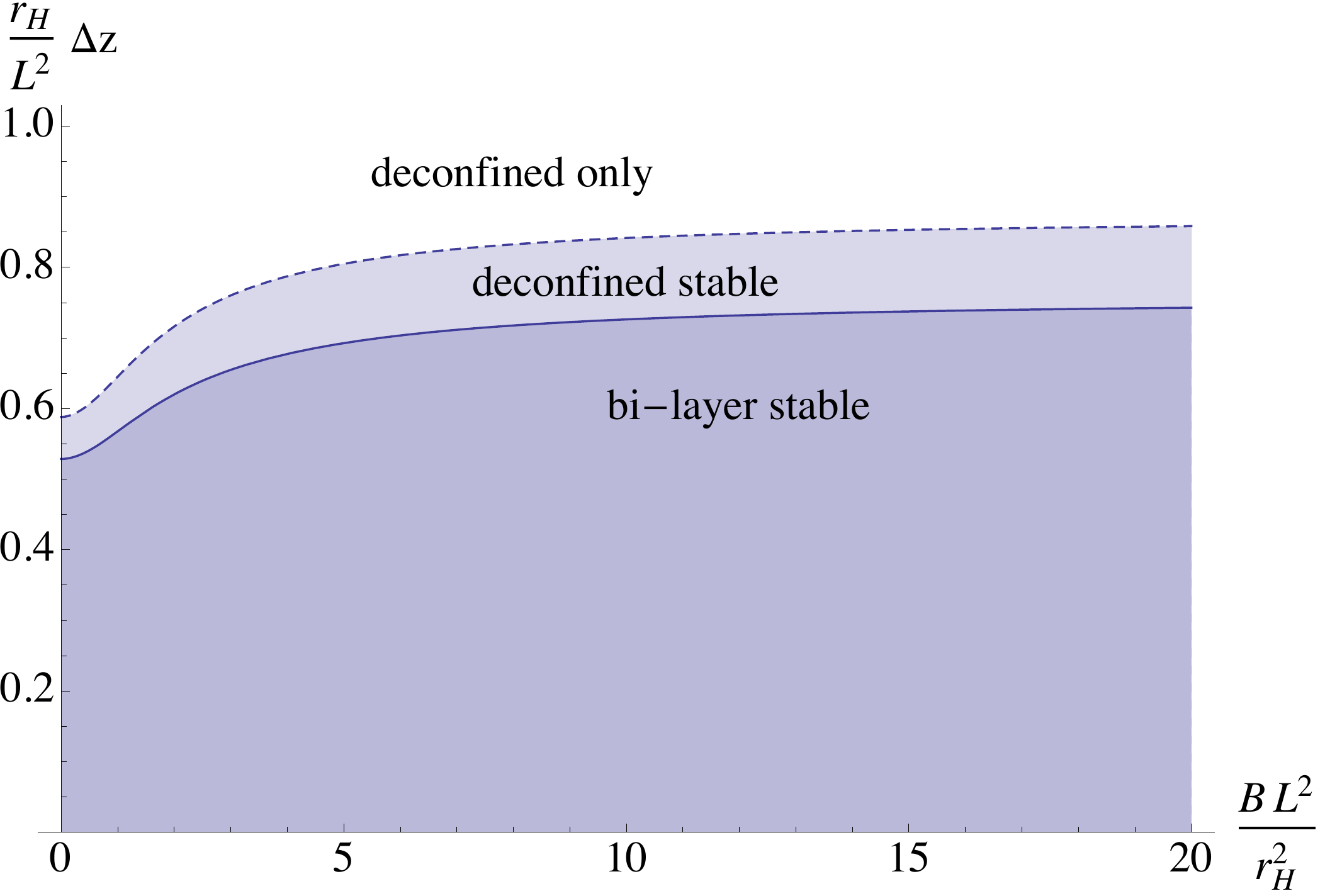}
   \caption{Phase diagram at non-zero temperature and external magnetic field for the single bilayer/ double monolayer (domain wall) phase transition.}
   \label{fig:phaseTBsingle}
\end{figure}

As before the stable phase is determined by the minimisation of the free energy. Substituting equations (\ref{U-TB}) and (\ref{parallel-TB}) into the wick rotated on-shell action and regularising as before, one finds the following expressions for the free energies of the U-shaped and parallel embeddings,
\begin{align}
F_{U}&= 2\,{\cal N'}\left(\,\int\limits_{r_z}^{\infty}dr\left(\,\frac{r^2\,(r^4+r_B^4)\,\sqrt{b(r)}}{\sqrt{r^4\,(r^4+r_B^4)\,b(r)-r_z^4\,(r_z^4+r_B^4)\,b(r_z)}}-r^2\right)-\frac{r_z^3}{3}\right) ,~~~\text{for}~r_z>r_H\ \label{F-U-T} \\
F_{||}&=2\,{\cal N'}\left(\,\int\limits_{r_H}^{\infty}dr\left(\sqrt{r^4+r_B^4}-r^2\right)-\frac{r_H^3}{3}\right)=-\frac{2\,{\cal N'}}{3}\,r_H^3\,{}_2F_1\left[-\frac{3}{4}\ ,-\frac{1}{2}\ ,\frac{1}{4}\ ,-\frac{r_B^4}{r_H^4}\right]\ . 
\end{align}
The numerical results can be summarised in the following phase diagram (figure \ref{fig:phaseTBsingle}). Note that the positive slope of the critical curve separating the deconfined (monolayer) and the bi-layer phases shows that, at fixed temperature (fixed $r_H$) and separation $\Delta z$ the bilayer phase is stabilised by the external magnetic field. This is yet another confirmation that magnetic catalysis is realised in this system.


\section{Meson spectrum of the single bilayer and stability analysis}
\label{sec:spectrum}

In this section, we will study the meson spectrum of the proposed model. It should be noted that the spectrum will be different from the one presented in \cite{Filev:2013vka}.
For the case at hand, the angle $\psi$ has a zero classical value, while there is a profile for the $x_3$ coordinate.   
Contrary to the approach followed in \cite{Filev:2013vka} (and also in \cite{Kuperstein:2008cq}), we will not introduce a new set of Cartesian coordinates
in order to realize even and odd boundary conditions. Instead, this will be achieved by a convenient reparametrization of the relevant equations in their  Schr\"odinger form. 

\noindent
We choose the following ansatz for the scalars,
\begin{eqnarray}
& z \, = \, z(r) \, + \,  (2\pi\alpha')\, \delta  z\left(t, r, \theta_+, \phi_- \right)\, , \quad 
\theta_{-} \, = \,  (2\pi\alpha')\,  \delta \theta_{m} \left(t, r, \theta_+, \phi_-\right) \, , &
\nonumber \\
&\phi_+ \, = \, \pi \, + \,  (2\pi\alpha')\,  \delta \phi_{p} \left(t, r, \theta_+, \phi_-\right) \, , \quad 
\psi  \,  = \,   (2\pi\alpha')\, \delta \psi\left(t, r, \theta_+, \phi_- \right)  \, , &
\end{eqnarray}
where the profile $z(r)$ is given by \eqref{profile-z}. 
The construction is supplemented with a $U(1)$ gauge field of the $D5$-brane, which contributes to quadratic order in the $\alpha'$ expansion.
Following closely the 
general prescription, we introduce the symmetric matrix $S$ in the following way, \footnote{Since, in this section, we are not studying the effect of the addition of the magnetic field, there is no antisymmetric contribution to the zeroth order expansion of the metric}
\begin{equation}
||{E_{ab}^0}||^{-1} \, = \, S \, ,
\end{equation}
while the non-zero elements are
\begin{eqnarray}
&& S^{tt} \, = \, G_{00}^{-1} \, ,  \quad  S^{11} \, = \, S^{22} \, = \, G_{11}^{-1} \, ,  \quad  S^{rr} \, = \, G_{rr}^{-1}  \, , 
\nonumber \\
&&
S^{++} \, = \, G_{\theta_+ \theta_+}^{-1} \, , \quad S^{--} \, = \, G_{\phi_- \phi_-}^{-1} \, ,  \label{S} 
\end{eqnarray}
with 
\begin{eqnarray}
&
G_{00} \, = \, g_{tt}^{(0)} \, ,  \quad 
G_{11} \, = \, g_{11}^{(0)} \, ,  \quad 
G_{rr} \, = \, g^{(0)}_{rr} \,+\, g^{(0)}_{33} \, z'(r)^2  \, ,& 
\nonumber\\
&
G_{\theta_+ \theta_+} \, = \, g^{(0)}_{\theta_+ \theta_+}  \, ,  \quad
G_{\phi_- \phi_-} \, = \, g^{(0)}_{\phi_- \phi_-} \,  . &
\end{eqnarray}
The non-cross terms in the quadratic expansion of the action are
\begin{eqnarray} \label{non_cross}
&&
- \, \frac{{\cal L}^{(2)}_{\delta \theta_m \delta \theta_m}}{\sqrt{-E_0}} \, = \, \frac{1}{2} \,
g^{(0)}_{\theta_- \theta_-}\, S^{ab}\partial_a\delta \theta_m \partial_b\delta \theta_m \, 
+ \, \left(\frac{1}{3} \, + \, \cot^2 \theta_{+} \right) \, \delta \theta_m^2 \,  , 
\nonumber\\
&&
-\, \frac{{\cal L}_{\delta \psi \delta \psi}^{(2)}}{\sqrt{-E_0}} \, = \, \frac{1}{2} \,  g^{(0)}_{\psi \psi}\, 
S^{ab}\partial_a\delta \psi\partial_b\delta \psi \, , 
\quad 
- \, \frac{{\cal L}_{\delta \phi_p \delta \phi_p}^{(2)}}{\sqrt{-E_0}} \, = \, \frac{1}{2} \, g^{(0)}_{\phi_+ \phi_+}\, 
S^{ab}\partial_a\delta  \phi_p \partial_b\delta  \phi_p  \,  ,
\\
&&
- \, \frac{{\cal L}_{\delta z \delta z}^{(2)}}{\sqrt{-E_0}} \, = \, \frac{1}{2} \, S^{rr} \,
S^{ab}\partial_a\delta  z \partial_b \delta  z  \, , 
\quad 
- \, \frac{{\cal L}_{\delta F \delta F}^{(2)}}{\sqrt{-E_0}} \, = \, \frac{1}{4} \,
S^{mp} \, S^{nq} \, F_{pq} \, F_{mn} \, ,
\nonumber
\end{eqnarray}
while the cross terms are
\begin{eqnarray} \label{cross}
&&
- \, \frac{{\cal L}_{\delta \phi_p \delta \psi}^{(2)}}{\sqrt{-E_0}} \, = \, g^{(0)}_{\phi_+ \psi}\, 
S^{ab}\partial_a\delta  \phi_p \partial_b\delta  \psi  \,  ,
\quad 
\frac{{\cal L}_{\delta \theta_m \delta \psi}^{(2)}}{\sqrt{-E_0}} \, = \, \frac{2}{3} \, \frac{1}{\sin \theta_{+}} \,
\delta \theta_m \partial_{\phi_-}\delta \psi 
\nonumber \\ 
&&
- \, \frac{{\cal L}_{\delta \theta_m \delta \phi_p}^{(2)}}{\sqrt{-E_0}} \, = \, \frac{2}{3}\, \cot \theta_{+}\, 
\delta \theta_m \partial_{\phi_-}\delta \phi_p \, .
\end{eqnarray}


\subsection{Fluctuation along $z$}

Since the scalar modes of $\delta z$ decouple from all the other modes, it is possible to solve them separately.
For this reason we apply the usual ansatz to separate variables 
\begin{equation} \label{z_ansatz}
\delta z \, = \, e^{i \omega t } \, h_3(r) \, \Theta(\theta_+) \, \Phi(\phi_- )
\end{equation}
and after redefining $r$ and $\omega$ as $r = \rho \,r_z$ and $\omega =  M\,r_z/L^2$, we have
\begin{eqnarray}
&&
h_3^{''}(\rho) \, + \, \frac{6}{\rho} \, \frac{\rho^8 \, + \, 1}{\rho^8 \, - \, 1}\, h_3^{'}(\rho) \, + \, \frac{3 \, \rho^6}{\rho^8 \, - \, 1}\, 
\left[ \kappa \, + \, \frac{M^2}{3 \, \rho^2}\right]\, h_3(\rho)  \,=\, 0
\label{EOMx3_radial} \\
&& 
\frac{\cot \theta_+  \,\Theta^{'}(\theta_+) }{\Theta(\theta_+)}\, + \, \frac{\Theta^{''}(\theta_+) }{\Theta(\theta_+)}\, 
+ \, \frac{1}{\sin^2 \theta_+} \, \frac{\Phi^{''}(\phi_-) }{\Phi(\phi_-)} \, = \, - \, \kappa \,  .
\label{EOMx3_angular}
\end{eqnarray}
Equation \eqref{EOMx3_angular} is easily recognized as the known differential equation for the two-sphere spherical harmonics
\begin{equation} \label{spherical_harmonics}
Y (\theta_+, \phi_-) \, \equiv \,\Theta (\theta_+) \, \Phi(\phi_-) \, = \, C_{l,m} \, P_l^m (\cos \theta_+) \, e^{i m \phi_-}  \quad 
\text{with} \quad 
\kappa \, = \, l \, \left(l \, + \, 1\right)\, 
\end{equation}
where $C_{l,m}$ is the normalization constant. 
The stability analysis of the fluctuations only requires the study of the lowest lying Kaluza-Klein mode, so from now on we set $\kappa = 0$ in \eqref{EOMx3_radial}.

\begin{figure}[h] 
   \centering
   \includegraphics[width=4in]{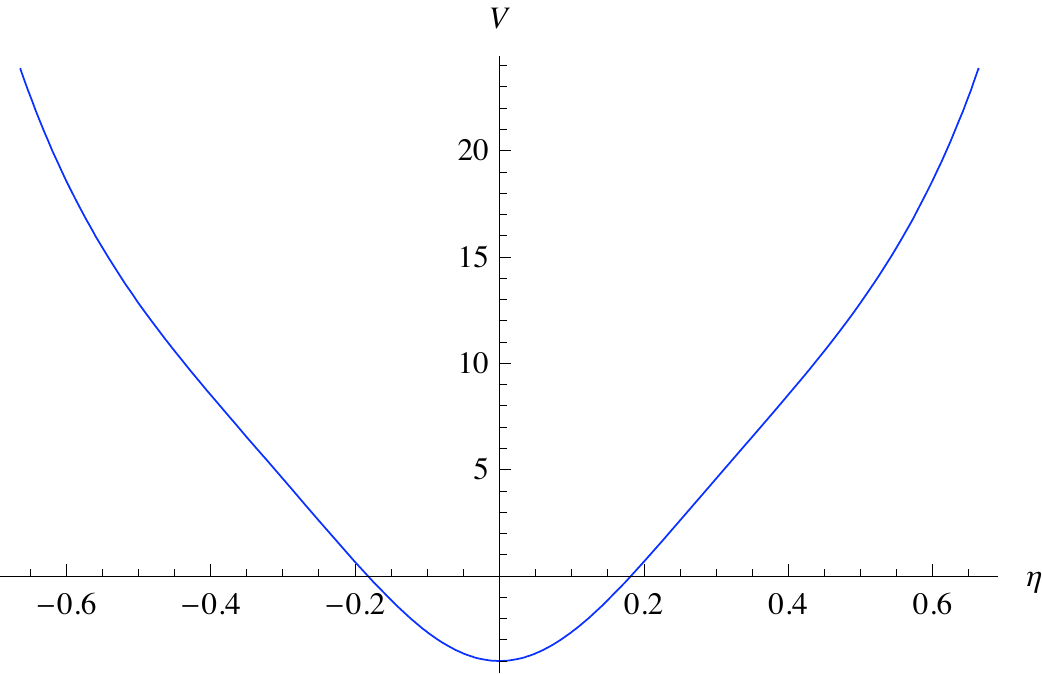} 
   \caption{Plot of the  Schr\"odinger potential as a function of $\eta$ in the case of fluctuations along $z$. The small negative part 
   near the region $\eta = 0$ is not sufficient to produce tachyonic modes. }
   \label{fig:potential}
\end{figure}

The coordinates that we considered cover only one branch of the U-shaped embeddings. To cover both branches and bring the equation of motion \eqref{EOMx3_radial} to 
a Schr\"odinger form we consider the following coordinate and functional change
\begin{equation} \label{schr-change}
\eta(\rho) \, =\pm\left( \,\frac{\sqrt{\pi} \,\Gamma(\frac{9}{8})}{\Gamma(\frac{5}{8})} \, - 
\, \frac{1}{\rho}\,{}_2F_1\left[\frac{1}{8}\ ,\frac{1}{2}\ ,\frac{9}{8}\ ,\frac{1}{\rho^8}\right]\right) \, \quad \mathrm{and} \quad 
\Psi(\eta) \, = \,\rho^2\,  h_3 \left(\rho(\eta)\right) \, \sqrt{1 \, - \frac{1}{\rho(\eta)^8}} \, , 
\end{equation}
where $\eta \in [-\frac{\sqrt{\pi} \,\Gamma(\frac{9}{8})}{\Gamma(\frac{5}{8})} ,\frac{\sqrt{\pi} \,\Gamma(\frac{9}{8})}{\Gamma(\frac{5}{8})}]$ and a different sign of $\eta$ corresponds to a different branch of the U-shaped embedding. The differential equation for the function $\Psi(\eta)$ is
\begin{equation}
\Psi^{''}(\eta) \, + \, \left(M^2 \, - \, V(\eta)  \right) \, \Psi(\eta) \,= \, 0 \quad \text{with} \quad V(\eta) \, = \, \frac{6 \,\rho(\eta)^8 \, - \, 10}{\rho(\eta)^6} \, . 
\end{equation}
Plotting the potential as a function of $\eta$, it is easy to notice (see figure \ref{fig:potential}) that for a wide range of parameter space the potential is positive. Nevertheless, a small negative part 
near the region $\eta = 0$ exists. Numerical computation of the spectrum of fluctuations will show that this is not sufficient to produce tachyonic modes 
and the spectrum is indeed tachyon free.

Solving \eqref{EOMx3_radial} approximately around $\rho = 1$ reveals two classes of possible solutions
\begin{equation} \label{2-classes}
h_3(\rho) \, \approx A \, + \,  \frac{B}{\sqrt{\rho\, - \, 1}} \, ,
\end{equation}
while solving the  Schr\"odinger equation around $\eta = 0$ perturbatively, it is clear that  there are two types of modes (as in \cite{Kuperstein:2008cq}), namely 
even and odd modes. Inverting the RHS of \eqref{schr-change}, it is possible to relate the two classes of \eqref{2-classes} with the even and odd types of modes. 
Doing this identification, we conclude that 
\begin{eqnarray}
&& A \, = \, 0 \quad \mathrm{and} \quad B \, \neq \, 0 \quad  \Rightarrow \quad  \Psi^{'}(0)\, = \, 0 
\quad  \mathrm{and} \quad \Psi(0) \, \neq \, 0 \quad  \Rightarrow \; \text{even modes} \label{even}
\\
&& A \, \neq \, 0 \quad  \mathrm{and} \quad B \, = \, 0 \quad  \Rightarrow \quad  \Psi^{'}(0)\, \neq \, 0 
\quad  \mathrm{and} \quad \Psi(0) \, = \, 0 \quad  \Rightarrow \; \text{odd modes} \label{odd}
\end{eqnarray}

We conclude this subsection by solving numerically \eqref{EOMx3_radial} after imposing either odd or even boundary conditions, in the fashion outlined above. 
For the first several excited states (in the $l=0$ case), we find
\begin{eqnarray}
&& M_{\rm even} \, = \, 2.323, \, 6.209, \, 9.086, \, 11.873, \,  \dots \\
&& M_{\rm odd} \, = \, 4.693, \, 7.668, \, 10.486, \, 13.252,  \, \dots 
\end{eqnarray}
As claimed above, the numerical computation confirms that the spectrum is tachyon free.


\subsection{Fluctuation along $\theta$}

Following the same strategy as in \cite{Filev:2013vka}, we suppress the $\phi_-$ dependence. This leads to a full decoupling of the $ \delta \theta_m$ modes from the rest of the modes. 
Applying the usual separation of  variables ansatz
\begin{equation}
 \delta \theta_m\, = \, e^{i \omega t } \, h(z) \, Y(\theta_+) \, .
\end{equation}
and redefining $r$  and $\omega$ as $r = \rho \,r_z$ and $\omega =  M\,r_z/L^2$, we have
\begin{eqnarray}
&&
h^{''}(\rho) \, + \, \frac{4 \, \rho^7}{\rho^8 \, - \, 1}\, h^{'}(\rho) \, + \, \frac{\rho^6}{\rho^8 \, - \, 1}\, 
\left[ \kappa \, + \, \frac{M^2}{\rho^2}\right]\, h(\rho)  \,=\, 0 \,  , 
\label{EOMh}\\
&&
Y^{''}(\theta_+) \, + \,\cot\theta_{+} \,Y^{'}(\theta_+) 
\, - \, {1 \over 3} \, \left( \kappa - 2 + {3 \over \sin^2 \theta_+} \right) Y(\theta_+) \, = \, 0 \,  .
\label{EOMY}
\end{eqnarray}
Treating \eqref{EOMY} as in \cite{Filev:2013vka}, it is possible to quantize $\kappa$ as follows
\begin{equation}
\kappa \,= \, - \, 4 \, - \, 3 \, m\left (m\, + \, 3 \right) \quad \text{with} \quad m \, > \, 0 \, .
\end{equation}
Since we are interested in the stabilty of the spectrum, we will focus on the 
 lowest lying Kaluza-Klein modes, implying $\kappa=-4$.

The  Schr\"odinger form of \eqref{EOMx3_radial} comes from the following coordinate  and functional change ,
\begin{equation} \label{schr-change}
\eta(\rho) \, =\pm \,\left(\,\frac{\sqrt{\pi} \,\Gamma(\frac{9}{8})}{\Gamma(\frac{5}{8})} \, - 
\, \frac{1}{\rho}\,{}_2F_1\left[\frac{1}{8}\ ,\frac{1}{2}\ ,\frac{9}{8}\ ,\frac{1}{\rho^8}\right]\, \right)\, \quad  \mathrm{and} \quad 
\Psi(\eta) \, = \,\rho \,  h_3 (\rho)  \, , 
\end{equation}
where again the different signs of $\eta$ correspond to the different branches of the U-shaped embedding. The differential equation for the function $\Psi(\eta)$ is given by:
\begin{equation}
\Psi^{''}(\eta) \, + \, \left(M^2 \, - \, V(\eta)  \right) \, \Psi(\eta) \,= \, 0 \quad 
\text{with} \quad V(\eta) \, = \, \frac{6 \,\rho(\eta)^8 \, + \, 2}{\rho(\eta)^6} \, > \, 0 \, . 
\end{equation}
The fact that the potential is strictly positive implies that the spectrum is tachyon free. 

As a final consistency check of the stability of the spectrum we solve numerically \eqref{EOMh} for $\kappa=-4$, 
imposing even and odd boundary conditions along the lines of \eqref{even} and \eqref{odd}. 
For the first several excited states, the result is 
\begin{eqnarray}
&& M_{\rm even} \, = \, 3.590, \, 6.465, \, 9.249, \, 11.994,  \, \dots\\
&& M_{\rm odd} \, = \, 5.025, \, 7.863, \, 10.624, \, 13.360, \, \dots
\end{eqnarray}


\subsection{Fluctuations along $\psi$ and $\phi$}

As can be seen from \eqref{cross}, after suppressing  the $\phi_-$ dependence only the fluctuations of $\delta \psi$ and $\delta\phi_p$ couple between themselves. 
Since we are interested in the lowest lying  Kaluza-Klein mode, we consider the following ansatz
\begin{equation}
 \delta \psi \, = \, e^{i \omega t } \, h_{\psi}(z) \, \cos\theta_+ \quad  \mathrm{and} \quad  \delta \phi_p \, = \, e^{i \omega t } \, h_{\phi}(z) \, .
\end{equation}
In this way, after redefining $r$ and $\omega$ as $r = \rho \,r_z$ and $\omega =  M\,r_z/L^2$, we end up with a coupled system of differential equations for $h_{\psi}$ and $h_{\phi}$
\begin{eqnarray}
&& h_{\psi}^{''}(\rho) \, + \, \frac{4 \, \rho^7}{\rho^8 \, - \, 1} \, h_{\psi}^{'}(\rho) \, + \, \frac{\rho^6}{\rho^8 \, - \, 1} 
\left[ \frac{M^2}{\rho^2} \, - \, 10 \right] \,  h_{\psi}(\rho) \, = \, 0 
\label{eqpsi}
\\
&& h_{\phi}^{''}(\rho) \, + \, \frac{4 \, \rho^7}{\rho^8 \, - \, 1} \, h_{\phi}^{'}(\rho) \, + \, \frac{\rho^4 \, M^2}{\rho^8 \, - \, 1} \,  h_{\phi}(\rho) \, +
\, \frac{2 \rho^6}{\rho^8 \, - \, 1} \, h_{\psi}(\rho) \, = \, 0 \, .
\label{eqphi}
\end{eqnarray}
Defining a new function $\Lambda(\rho)$ as follows, 
\begin{equation} \label{def-L}
\Lambda(\rho)\, = \, h_{\psi}(\rho) \, + \, 5 \, h_{\phi}(\rho) \,  ,
\end{equation}
it is possible to diagonalize the above system of coupled differential equations 
\begin{equation}
\Lambda^{''}(\rho) \, + \, \frac{4 \, \rho^7}{\rho^8 \, - \, 1} \, \Lambda^{'}(\rho) \, + \, \frac{\rho^4 \, M^2}{\rho^8 \, - \, 1}  \,  \Lambda(\rho) \, = \, 0 \, .
\label{eqL}
\end{equation}

The  Schr\"odinger form of \eqref{eqpsi} and \eqref{eqL} comes from the same coordinate and functional change we used for the $\theta$ fluctuations, namely \eqref{schr-change}.
The forms of the potentials in this case are given by 
\begin{equation}
 V_{\psi}(\eta) \, = \, 2 \, \frac{6 \,\rho(\eta)^8 \, + \, 2}{\rho(\eta)^6} \, > \, 0 
\quad  \mathrm{and} \quad 
V_{\Lambda}(\eta) \, = \, \frac{2 \,\rho(\eta)^8 \, + \, 2}{\rho(\eta)^6} \, > \, 0 \, . 
\end{equation}
As previously, a strictly positive potential implies a spectrum without tachyonic modes, and as a consistency check we again numerically evaluate, using even and odd boundary conditions,
the first several excited states both for $\psi$,
\begin{eqnarray}
&& M^{\psi}_{\rm even} \, = \, 4.578, \, 7.578, \, 10.419, \, 13.199,  \, \dots\\
&& M^{\psi}_{\rm odd} \, = \, 6.099, \, 9.010, \, 11.814, \, 14.577, \, \dots
\end{eqnarray}
and $\Lambda$,
\begin{eqnarray}
&& M^{\Lambda}_{\rm even} \, = \, 2.606, \, 5.320, \, 8.043, \, 10.760,  \, \dots\\
&& M^{\Lambda}_{\rm odd} \, = \, 3.918, \, 6.680, \, 9.400, \, 12.111, \, \dots
\end{eqnarray}
Note also that for $M=0$ equation \eqref{eqL} has the special solution $\Lambda = const$. Similarly to the situation in \cite{Kuperstein:2008cq, Filev:2013vka},
this Goldstone mode corresponds to the spontaneously broken conformal symmetry.


\subsection{Fluctuations along the worldvolume gauge fields}

Following the analysis of \cite{Kuperstein:2008cq,  Filev:2013vka}, we will focus our attention on the modes independent of the two-sphere coordinates, and also freezing the dependence on 
$\theta_{+}$ and $\phi_-$, allowing coordinate dependence only along $t, x_1, x_2$ and $r$. 
We will also ignore all the components of the gauge field along the $\theta_{+}$ and $\phi_-$ directions. The main lesson of the equivalent analysis in \cite{Filev:2013vka} was the 
presence of two Goldstone modes, one scalar and one vector, both of them are renormalizable. This feature, which cannot be observed in the model of \cite{Kuperstein:2008cq}, continues
to exist also in the present analysis of the fluctuations of the gauge fields of the single bilayer. The detailed analysis of the spectrum will be presented in the following.

The reduced action for the fluctuations of the gauge field is
\begin{equation}
S \, = \, - \, (2\pi\alpha')^2 \, {\cal N} \, \int\,d^3x\,dr\,\Big[ \, C(r)\,F_{\mu\nu}F^{\mu\nu}\, + \, 2 \, D(r) \, F_{\mu r} \, F^{\mu~}_{~r} \, \Big]\,
\end{equation}
where
\begin{equation}
C(r) \, = \, \frac{\pi\,L^4}{3} \, \frac{r^2}{\sqrt{r^8 \, - \, r_z^8}}
\quad  \mathrm{and} \quad 
D(r) \, = \, \frac{\pi}{6 \, r^2}\,\sqrt{r^8 \, - \, r_z^8}\ .
\end{equation}
Changing the radial coordinate from $r$ to $\xi$ as follows,
\begin{equation}
\xi(r) \, = \, \pm \, \frac{1}{\sqrt{2}} \, \int\limits_{r_z}^r dz \, \sqrt{\frac{C(z)}{D(z)}}\,= \, \pm \, L^2 \, \Bigg[ \frac{\sqrt{\pi}}{r_z}\frac{\Gamma(\frac{9}{8})}{\Gamma(\frac{5}{8})} \, - 
\, \frac{1}{r}\,{}_2F_1\left[\frac{1}{8}\ ,\frac{1}{2}\ ,\frac{9}{8}\ ,\frac{r_z^8}{r^8}\right] \Bigg] \, ,
\end{equation}
where we need both signs of $\xi$ in order to cover the two branches of the $D5$-brane and the $\overline{D5}$-brane
, we arrive at
\begin{equation}
S \, = \, - \, T'\,\int\,d^3x\,\int\limits_{-\xi_*}^{\xi_*}\,d\xi \, \left(\frac{1}{4} \, F_{\mu\nu} \, F^{\mu\nu} \, + \, \frac{1}{2} \,  F_{\mu\xi} \, F^{\mu~}_{~\xi}\right)\ ,
\end{equation}
where
\begin{equation}\label{Sff-red}
T' \, = \, \frac{4}{3}\,\pi\,L^2\,(2\pi\alpha')^2{\cal N} \quad \text{and} \quad 
\xi_* \, = \, \frac{L^2 \, \sqrt{\pi}}{r_z}\,\frac{\Gamma\left(\frac{9}{8}\right)}{\Gamma\left(\frac{5}{8}\right)}\ .
\end{equation}
Next, again following closely the analysis of  \cite{Filev:2013vka}, we proceed to expand the components of the gauge field as
\begin{equation}
A_{\mu}(x,\xi)\,=\,\sum\limits_n\,a_{\mu}^{n}(x)\,\alpha^{n}(\xi) \quad  \mathrm{and} \quad 
A_{\xi}(x,\xi)\,=\,\sum\limits_n\,b^{n}(x)\,\beta^{n}(\xi)\ .
\end{equation}
The functions $\alpha^n$ are defined in the interval $\xi \in [-\xi_*,\xi_*]$ and a convenient choice of basis turns out to be 
\begin{equation}\label{basisalpha}
\alpha^n \, = \, \frac{1}{\xi_*^{1/2}}\,\cos (M_n\,\xi) \quad \text{with} \quad 
M_n \, = \, \frac{n\,\pi}{2 \, \xi_*} \, = \, \frac{\sqrt{\pi}}{2} \, \frac{\Gamma\left(\frac{5}{8}\right)}{\Gamma\left(\frac{9}{8}\right)}\,n  
\end{equation}
We find that the zero mode $\alpha_0=const$ is normalizable, as in \cite{Filev:2013vka}, . 
A convenient choice of basis to parametrize the functions $\beta^n$ is 
\begin{equation}
\beta^n \, = \,     \begin{cases} \frac{1}{M_n}\,\partial_{\xi}\alpha^n \, = \, - \, \frac{1}{\xi_*^{1/2}}\,\sin(M_n\,\xi) & \text{for}~n\geq1\\ 
\alpha_0 \, = \, \frac{1}{\xi_*^{1/2}} & \text{for}~n=0 \end{cases}
\label{basisbeta}\, .
\end{equation}
With this choice of basis for the functions $\alpha^n$ and $\beta^n$, and a gauge transformation 
$a_{\mu}^n\to a_{\mu}^n+\frac{1}{M_n}\partial_{\mu}b^n$ (for $n\geq1$),  the total action for the meson modes becomes 
\begin{equation}
S=-T'\int\,d^3x\,\left\{\frac{1}{2}\partial_{\mu}b^0\,\partial^{\mu}b^0+\frac{1}{4}f_{\mu\nu}^0\,f^{\mu\nu\,0}+\sum\limits_{n=1}^{\infty}\left[\frac{1}{4}f_{\mu\nu}^n\,f^{\mu\nu\,n}+\frac{1}{2}M_n^2\,a_{\mu}^n\,a^{\mu\,n}\right] \right\}\ ,
\end{equation}
where $M_n$ is given by \eqref{basisalpha}. 

The spectrum of the fluctuations of the gauge field gives rise to a plethora of fields: massive (for $n\geq1$) and massless  (the $n=0$ mode) vector fields
as well as a massless scalar field $b^0$. This latter mode  
is associated with the Goldstone mode of the spontaneously broken $U(1)\times U(1)$ chiral symmetry, again in complete analogy with the analysis in \cite{Kuperstein:2008cq}.

\section{Combined bilayer and monolayer phase}
\label{sec:combined}

\begin{figure}[h] 
   \centering
   \includegraphics[width=4in]{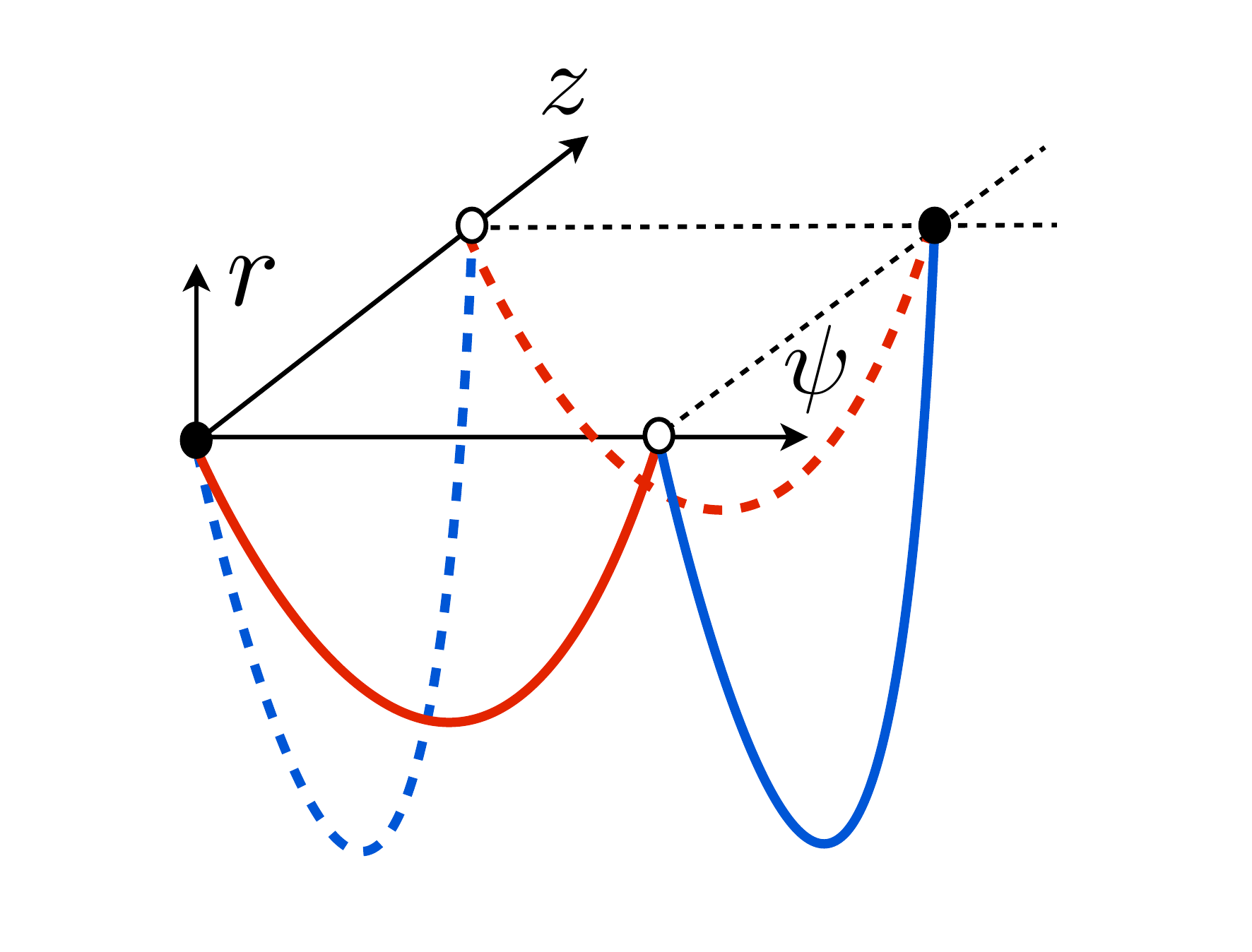} 
   \caption{Schematic of possible flavour brane configurations in the $\left(x^3, \psi, r\right)$-subspace. The filled (empty) circles indicate orientation corresponding to $D5$($\overline{D5}$)-branes, so the U-shaped embeddings have to connect a filled and an empty dot.
   The red curves (monolayer phase) correspond to two copies of the $\psi(r)$ embeddings studied in \cite{Filev:2013vka}, while the blue curves (bilayer phase) represent two (2+1)-dimensional layers which are separated (in the $z$ direction) in the UV and connect smoothly in the IR. In principle, it would also be possible to find some "diagonal" embedding along a linear combination $v_\pm := z \pm \frac{L^2}{3r} \psi$. It is intuitively clear, however, that these embeddings should have higher free energies than the extremal embeddings and therefore we do not consider them here.}
   \label{fig:schematic}
\end{figure}

In this section we are going to explore in detail the different possible flavour brane configurations in our framework. Besides studying the embedding equations, this includes a thorough investigation of the free energies and 
phase transitions at finite temperature and external magnetic field.
Figure \ref{fig:schematic} schematically shows possible flavour brane embeddings in the $\left(x^3,\psi,r\right)$-submanifold of the background. 


\subsection{Zero temperature}

The natural starting point for our investigation is the case of zero temperature and without exciting any external fields. From the induced metric on the $D5/\overline{D5}$-branes,
\begin{align}\label{eq:D5metric}
ds^2=\frac{r^2}{L^2}\left(-dt^2+dx_1^2+dx_2^2\right)+\frac{L^2}{r^2}\left[dr^2\left(1+\frac{r^4}{L^4}z'(r)^2+\frac{r^2}{9}\psi'(r)^2\right)+\frac{r^2}{3}d\Omega_2^2 \right],
\end{align}
we obtain the relevant flavour brane DBI action,
\begin{equation}\label{eq:DBI}
S_{\mathrm{D5}}=- \tau_5 \int d \xi^6 \sqrt{\det P[g]}=- 2 \mathcal{N} \int dr \, r^2 \sqrt{1+\frac{r^4}{L^4}z'(r)^2+\frac{r^2}{9}\psi'(r)^2},
\end{equation}
where $\mathcal{N}= \frac{2 \pi}{3}\tau_5 \mathrm{Vol}(\mathbb{R}^{2,1})$. 
From this, we can straightforwardly deduce the equations of motion which can be integrated once, due to the cyclic nature of the $z$ and $\psi$ coordinates,
\begin{equation} \label{eqMotPzps}
\frac{\frac{r^6}{L^4}z'(r)}{\sqrt{1+\frac{r^4}{L^4}z'(r)^2+\frac{r^2}{9}\psi'(r)^2}}= \Pi_z  \quad \text{and} \quad  
\frac{\frac{r^4}{9}\psi'(r)}{\sqrt{1+\frac{r^4}{L^4}z'(r)^2+\frac{r^2}{9}\psi'(r)^2}}= \Pi_{\psi},
\end{equation}
where we introduced the conserved quantities and canonically conjugate momenta $\Pi_z$ and $\Pi_{\psi}$. Solving for $z'(r)$ and $\psi'(r)$ we obtain
\begin{equation} \label{z_psi_solsm}
z'(r)=\pm\,\frac{\frac{L^4}{r^2}\,\Pi_z}{\sqrt{r^8-L^4\,\Pi_z^2-9\,r^2\,\Pi_{\psi}^2}}  \quad \text{and} \quad  
\psi'(r)=\pm\,\frac{9\,\Pi_\psi}{\sqrt{r^8-L^4\,\Pi_z^2-9\,r^2\,\Pi_{\psi}^2}} \, .
\end{equation}
We see that equations in \eqref{z_psi_solsm} describe two branches of the D5-brane embedding. For real values of $\Pi_z$ and $\Pi_\psi$, 
there is a minimal radial distance $r_0$ at which the denominators in  \eqref{z_psi_solsm} vanish and the two branches of the D-brane join smoothly to form a U-shaped embedding. 
Only when both conjugate momenta vanish ($\Pi_z=\Pi_\psi=0$), the solution corresponds to straight embeddings localized at constant $z$ and $\psi$.

We consider the following configurations at zero temperature:
\begin{itemize}
\item The bilayer phase, i.e., the U-shaped embedding in the $z$-direction, for $\Pi_{\psi}=0$. This configuration was discussed in section \ref{sec:singlebilayer}.
\item The two monolayers phase, i.e., the U-shaped embedding in the $\psi$-direction, for $\Pi_z=0$. This was the subject of a recent paper by the present authors \cite{Filev:2013vka}.
The embedding equation and the analytic solution are in this case 
\begin{equation}
\psi'(r)= \frac{\Pi_{\psi}}{\sqrt{\frac{r^{8}}{81}-\frac{r^2}{9}\Pi_{\psi}^2}}=\frac{3 r_{\psi}}{r\sqrt{r^6-r_{\psi}^6}} \quad \text{with} \quad 
\psi(r)= \arccos \left(\frac{r_{\psi}^3}{r^3} \right) \, ,
\end{equation}
where $r_{\psi}$ represents the minimal radial position of the embedding. 
\end{itemize}

\subsection{Finite temperature and external magnetic field}

Let us introduce finite temperature through the inclusion of an emblackening factor $b(r):=1-\frac{r_H^4}{r^4}$ in the metric, and also turn on a U(1) gauge field on the probe branes in order to excite a magnetic field. 
Here, we will work with the ansatz $A_2 ={H} x^1$, which corresponds to a
constant magnetic field $F_{12} =H$ along the $x^3$ direction, i.e., perpendicular to the (2+1)-dimensional defects.
Taking into account the combined effects of finite temperature and constant magnetic field yields DBI action for the flavour $D5$ branes,
\begin{equation}\label{eq:DBI_TB}
S_{\mathrm{D5}} = - 2\,{\cal N}_T \int dr \, r^2  \sqrt{1 + B^2 \frac{L^4}{r^4}}\sqrt{1 + \frac{r^4b(r)}{L^4}z'(r)^2+ \frac{r^2}{9} b(r) \psi'(r)^2},
\end{equation}
where $B:= 2 \pi \alpha' H$. Thus, the resulting equation of motion reads
\begin{equation}
\frac{\frac{r^6}{L^4} \sqrt{1 + B^2 \frac{L^4}{r^4}} b(r) z'(r)}{\sqrt{1+\frac{r^4b(r)}{L^4}z'(r)^2+\frac{r^2 b(r)}{9}\psi'(r)^2}}= \Pi^{T,B}_{z} \quad \text{and} \quad 
\frac{\frac{r^4}{9}\sqrt{1 + B^2 \frac{L^4}{r^4}} b(r) \psi'(r)}{\sqrt{1+\frac{r^4b(r)}{L^4}z'(r)^2+\frac{r^2 b(r)}{9}\psi'(r)^2}}= \Pi^{T,B}_{\psi} \, ,
\end{equation}
where now $\Pi^{T,B}_z:= \frac{r_z^4}{L^2}\sqrt{b(r_z)}\sqrt{1+B^2\frac{L^4}{r_z^4}}$ and $\Pi^{T,B}_{\psi}:= \frac{r_{\psi}^3}{3}\sqrt{b(r_{\psi})}\sqrt{1+B^2\frac{L^4}{r_{\psi}^4}}$.
As before, we can again distinguish between two 'extremal' cases.


\subsubsection{Bilayer phase}
When $\Pi^{T,B}_{\psi}=0=\psi'(r)$, we find 
\begin{equation}
z'(r)=\pm\, \frac{L^2 r_z^4\sqrt{b(r_z)}\sqrt{1+B^2 \frac{L^4}{r_z^4}}}{r^2 \sqrt{r^8b^2(r)\left(1+B^2\frac{L^4}{r^4}\right)-r_z^8b(r)b(r_z)\left(1+B^2\frac{L^4}{r_z^4}\right)}}.
\label{prprp}
\end{equation}
Introducing $r_B:=BL^2$, and for $T=0$ and thus $b(r)=1$, namely the case that we will be mostly interested in subsequently, this simplifies to
\begin{equation}
z'(r)= \pm\,\frac{L^2 r_z^2 \sqrt{r_z^4+r_B^4}}{r^2 \sqrt{r^8+r_B^4\left(r^4-r_z^4\right) -r_z^8}}.
\end{equation} 
The analysis of this case is identical to the analysis of the single bilayer presented in section \ref{SBLTB}. 
The solution to equation (\ref{prprp}) is thus given by \eqref{U-TB}.


\subsubsection{Monolayer phase}
On the other hand, setting $\Pi^{T,B}_{z}=0=z'(r)$, we arrive at the familiar case \cite{Filev:2013vka},
\begin{equation}
\psi'(r)=\pm\,\frac{3 r_{\psi}^3 \sqrt{b(r_{\psi})}\sqrt{1+B^2\frac{L^4}{r_{\psi}^4}}}{r\sqrt{r^6b^2(r)\left(1+B^2\frac{L^4}{r^4}\right)-r_{\psi}^6 b(r)b(r_{\psi}) \left( 1+B^2\frac{L^4}{r_{\psi}^4}\right)}},
\end{equation}
which reduces for zero temperature but finite magnetic field to 
\begin{equation}
\psi'(r)=\pm\,\frac{3 r_{\psi} \sqrt{r_{\psi}^4+r_B^4}}{r\sqrt{r^6+r_B^4\left(r^2-r_{\psi}^2\right)-r_{\psi}^6}} \, .
\end{equation}
This can be integrated to yield the numeric result for the $\psi(r)$ embedding (for a detailed discussion, cf. \cite{Filev:2013vka}).
%


\subsection{Phase structure}
In order to be able to decide which phase is energetically favoured, we have to evaluate and compare the (regularised) free energy densities of the different possible phases.
\subsubsection{Zero temperature}
The regularised free energy density of the U-shaped monolayer phase can be expressed as
\begin{equation}
F_{U_{\psi}}= 2\,{\cal N'}\,\int_{r_{\psi}}^{\infty} dr \left( \frac{r^5}{\sqrt{r^6-r_{\psi}^6}}- {r^2} \right) - \frac{r_{\psi}^3}{3}\, = \, 0 \, ,
\end{equation} 
while the regularised free energy density of the U-shaped bilayer phase is given by
\begin{equation}
F_{U_z}=2\,{\cal N'}\, \int_{r_{z}}^{\infty} dr\left( \frac{r^6}{ \sqrt{r^8-r_{\psi}^8}}- r^2 \right) - \frac{r_{z}^3}{3} \, = \,  
\frac{{\cal N'}\,\sqrt{\pi}\, \Gamma\left(-\frac{3}{8}\right) \,  r_z^3}{4 \,\Gamma\left(\frac{1}{8}\right)}\,<\,0,
\end{equation}
which, for $r_z >0$, is always negative. Therefore, $\Delta F = F_{U_{z}}-F_{U_{\psi}} <0$, and the bilayer phase is energetically favoured at zero temperature. 
\subsubsection{Zero temperature and finite magnetic field}
The regularised free energies are now given by 
\begin{align}
F_{U_{\psi}}&= 2\,{\cal N'}\,\int_{r_{\psi}}^{\infty} dr \left( \frac{r(r^4+r_B^4)}{ \sqrt{r^2(r^4+r_B^4)-r_{\psi}^2(r_{\psi}^4+r_B^4)}}- {r^2} \right) - \frac{r_{\psi}^3}{3},\\
F_{U_z}&=2\,{\cal N'}\, \int_{r_{z}}^{\infty} dr \left( \frac{r^2(r^4+r_B^4)}{\sqrt{r^4(r^4+r_B^4)-r_{z}^4(r_{z}^4+r_B^4)}}- {r^2} \right) - \frac{r_{z}^3}{3}.
\end{align} 

Important physical quantities, from the dual field theory point of view, are the asymptotic separations of the brane and anti-brane in the UV, given by the following expressions,
\begin{align}
\Delta \psi_{\infty}&= 2 \int_{r_{0,\psi}}^{\infty} dr \frac{3 r_{\psi}^3}{r \sqrt{r^6+r^2 r_B^4-r_{\psi}^6}}, \quad \mathrm{Monolayer}\; \mathrm{phase}\\
\Delta z &= 2 \int_{r_{0,z}}^{\infty} dr \frac{r_z^4}{r^2 \sqrt{r^8+r^4 r_B^4-r_z^8}}, \quad \mathrm{Bilayer}\; \mathrm{phase}.
\end{align}
which we can evaluate numerically or find approximations.

As it was shown in \cite{Filev:2013vka}, the parameter $\Delta\psi$ ranges from zero to $3\pi$. However, since the length of the $\psi$-cycle is $4\pi$, at  $\Delta\psi=2\pi$ the two branches of the U-shaped embeddings are at antipodal points. Increasing the separation further to $\Delta\psi>2\pi$ is equivalent to having a separation $(4\pi-\Delta\psi)<2\pi$. Therefore, we will restrict the possible values of $\Delta\psi$ to the interval $[\,0\,,\,2\pi\,]$ and when more than one U-shaped embedding exists, the one with lower free energy will be selected. 

\begin{figure}[h] 
   \centering
     \includegraphics[width=5.0in]{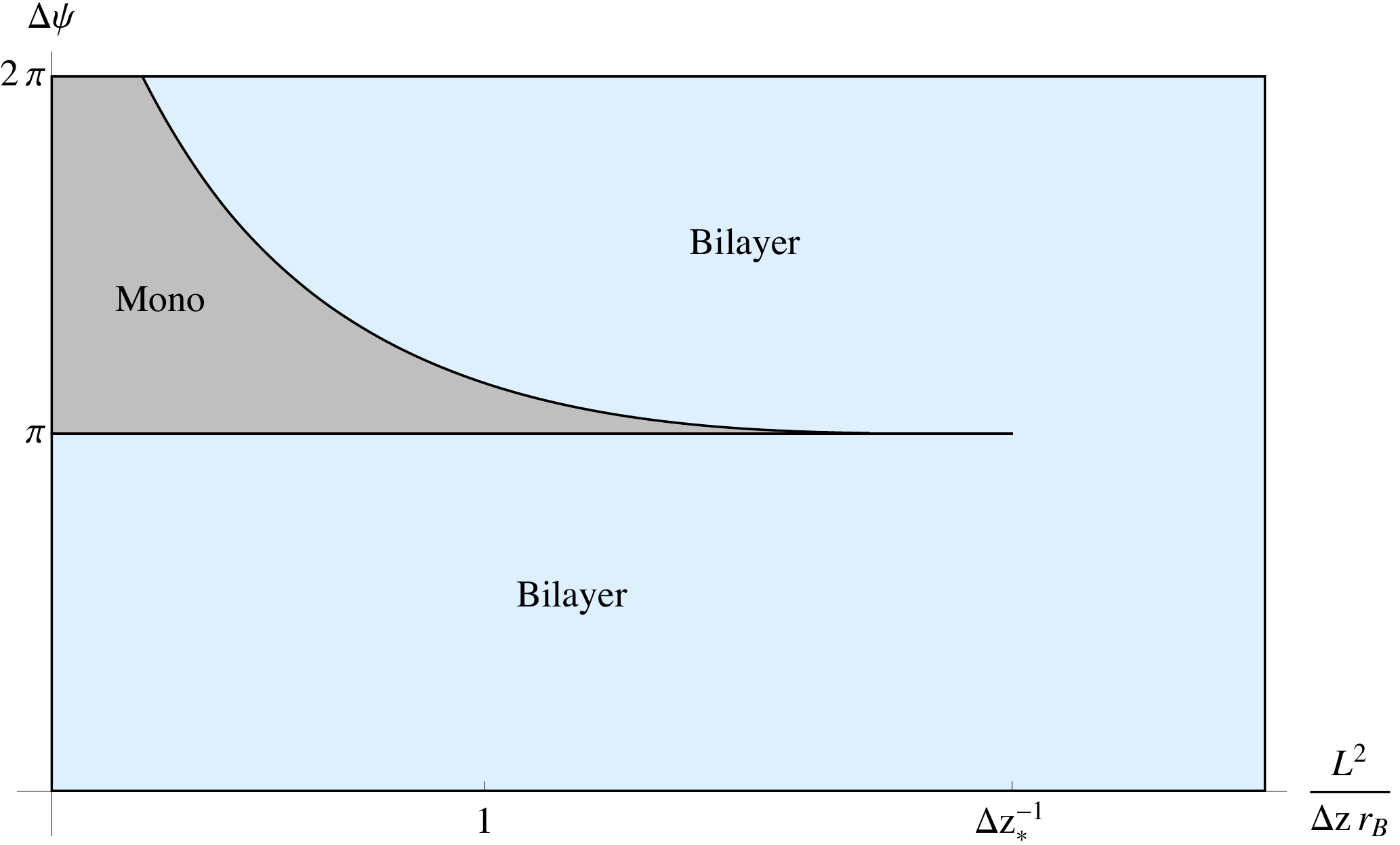} 
   \caption{Phase diagram at $T=0$ and finite external magnetic field.}
   \label{fig:phaseT0Bfin}
\end{figure}

If we compare the resulting free energies for the two possible U-shaped configurations, we find the following phase diagram presented in fig. \ref{fig:phaseT0Bfin}. As one can see form the figure, the mono layer phase exists only for $\pi<\Delta\psi <2\,\pi$ and $\Delta z >\Delta z_*$, where $\Delta z_*$ is determined by the zero of the regularised free energy $F_{U_z}$. 
\subsubsection{Finite temperature and finite magnetic field}
\begin{figure}[h] 
   \centering
   \includegraphics[width=2.75in]{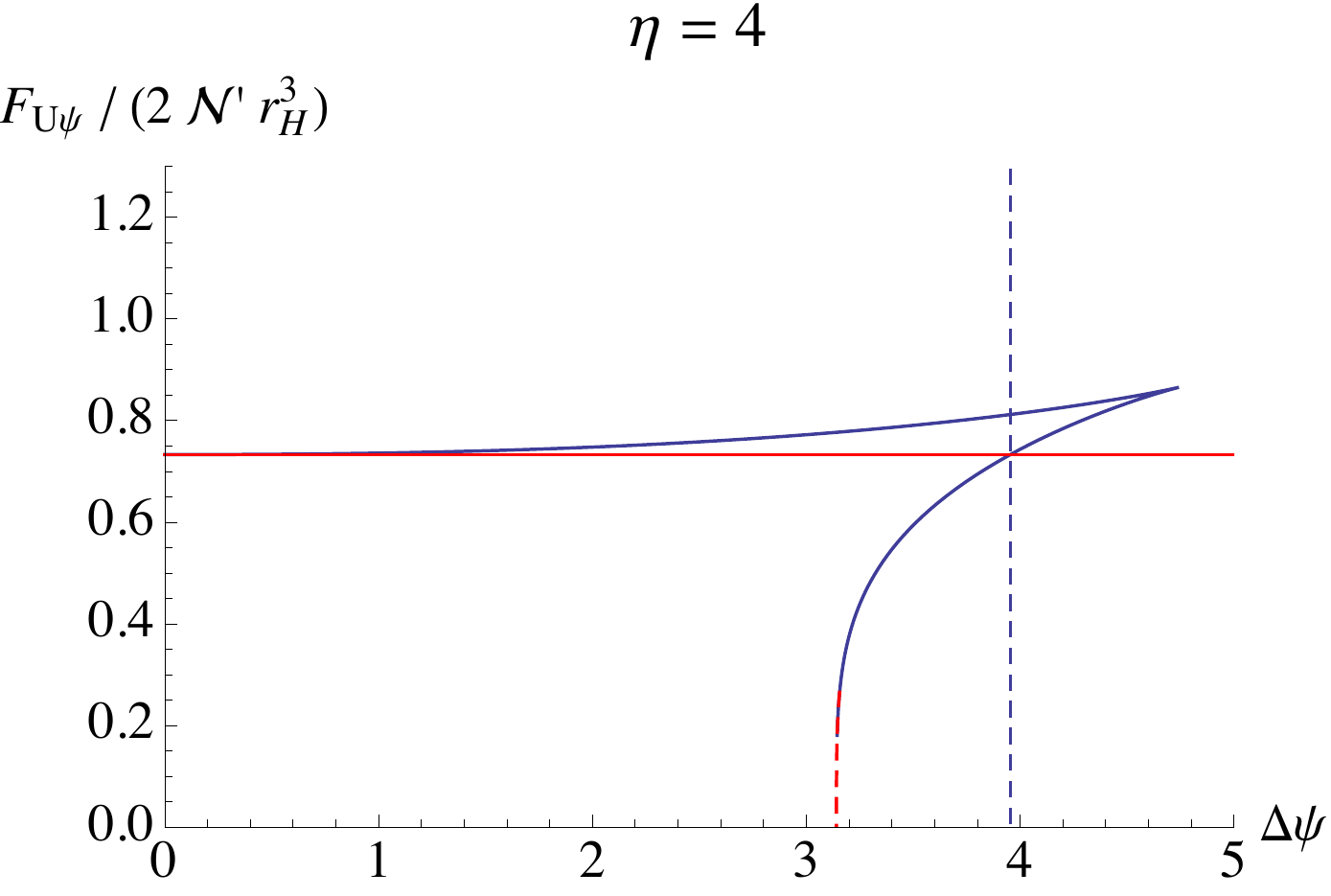}
    \includegraphics[width=2.75in]{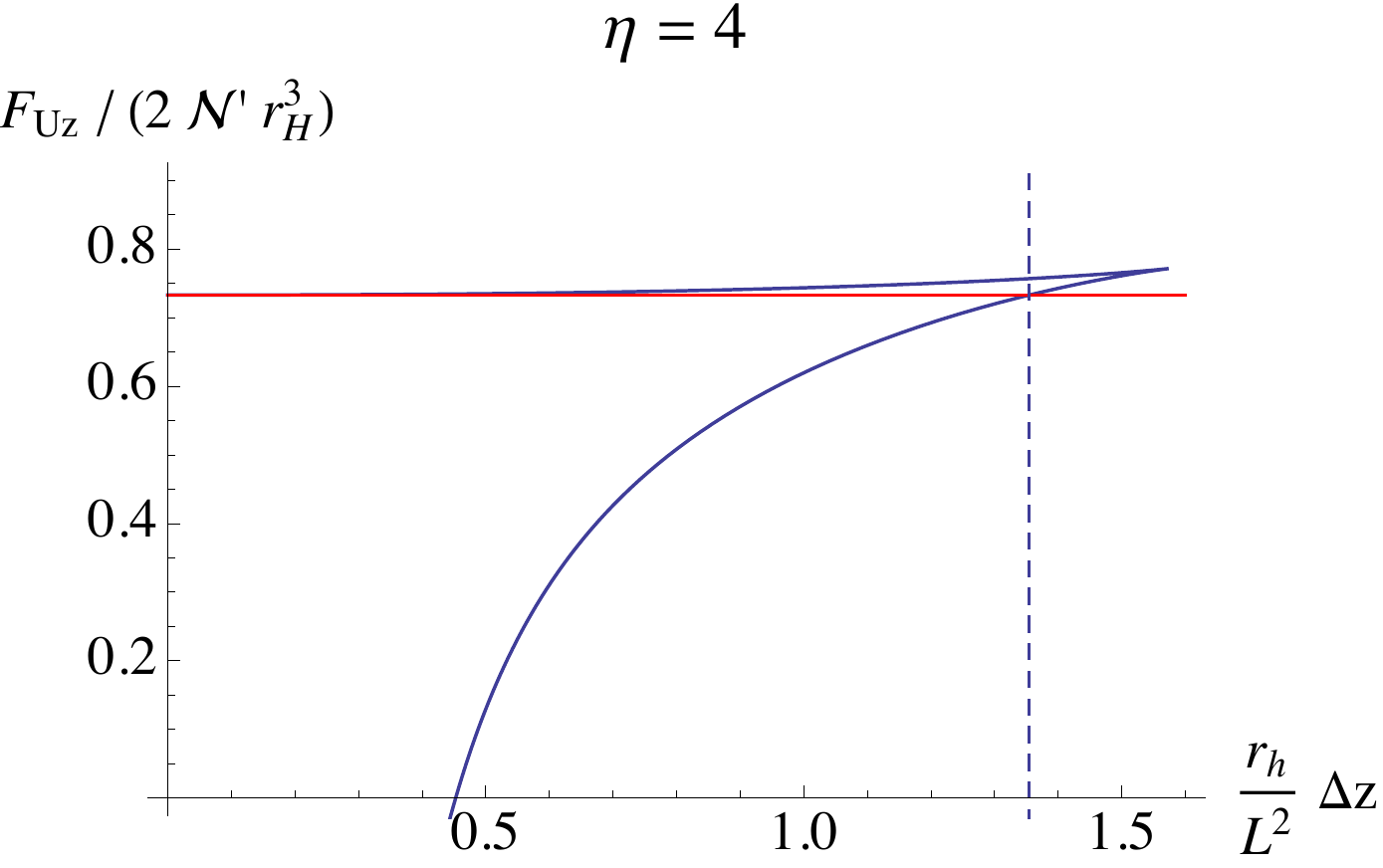}
  \caption{Free energies of the U-shaped (blue) and straight embeddings (red) for $\eta =4$. The vertical dashed lines represent the critical values for $\Delta \psi$ and $\Delta z$ respectively, for which the free energy of the U-shape embeddings and the straight embeddings are equal. The red dashed line in the left panel shows an approximate solution. }
   \label{fig:FUTB}
\end{figure}

In the following we will present our results for the phase structure of the bilayer and monolayer phases in the general case at finite temperature and finite external magnetic field. As before, it is advantageous to work with the rescaled variables,
\begin{equation}
\tilde{r}= \frac{r}{r_H},~~~ \; \widetilde{\Pi}^{T,B}_{\psi}= \frac{\Pi^{T,B}_{\psi}}{r_H^3},\;~~ ~ \widetilde{\Pi}^{T,B}_{z}= \frac{\Pi^{T,B}_{z}}{r_H^2}, \, \ldots
\end{equation}
In terms of these rescaled coordinates the regularised free energies are given by the following expressions, cf. eq.~(\ref{F-U-T}) and \cite{Filev:2013vka}, eqs.~(4.16) and (4.17),\footnote{Note the factor of $2$ difference compared to the cited equations.}
\begin{eqnarray}
\tilde F_{U_z}= \frac{F_{U_z}}{({2\cal N'}r_H^3)}&=&\int\limits_{\tilde r_{z}}^{\infty}\,d\tilde r\, \left(\frac{   \tilde r^2 \left(\tilde r^4+\eta^2\right)\sqrt{b(\tilde r)}  }{\sqrt{\tilde r^4\left(\tilde r^4+\eta^2\right) b(\tilde r)-\tilde r_{z}^4\left(\tilde r_{z}^4+\eta^2\right) b(\tilde r_{\psi})} }-\tilde r^2\,     \right)-\frac{\tilde r_{z}^3}{3} \, ,  \\
\tilde F_{U_{\psi}}= \frac{F_{U_{\psi}}}{({2\cal N'}r_H^3)}&=&\int\limits_{\tilde r_{\psi}}^{\infty}\,d\tilde r\, \left(\frac{   \tilde r \left(\tilde r^4+\eta^2\right)\sqrt{b(\tilde r)}  }{\sqrt{\tilde r^2\left(\tilde r^4+\eta^2\right) b(\tilde r)-\tilde r_{\psi}^2\left(\tilde r_{\psi}^4+\eta^2\right) b(\tilde r_{\psi})} }-\tilde r^2\,     \right)-\frac{\tilde r_{\psi}^3}{3} \, , \quad \quad \\
\tilde F_{||}=\frac{ F_{||}}{({2{\cal N'}}r_H^3)}&=&\int\limits_1^{\infty}\,d\tilde r (\sqrt{\tilde r^4+\eta^2}-\tilde r^2)-\frac{1}{3}=-\frac{1}{3}\,_2F_1\left(-\frac{3}{4},-\frac{1}{2},\frac{1}{4},-\eta^2\right)\, .
\label{ParallelF}
\end{eqnarray}
which can only be evaluated and compared numerically. \\
For any given $\eta$ one can compute numerically the free energies of the U-shaped monolayers and bilayers and compare it to the straight embeddings. An example is presented in figure \ref{fig:FUTB}  for $\eta=4$.

\begin{figure}[h] 
\hspace{-.9cm}      \includegraphics[width=3.4in]{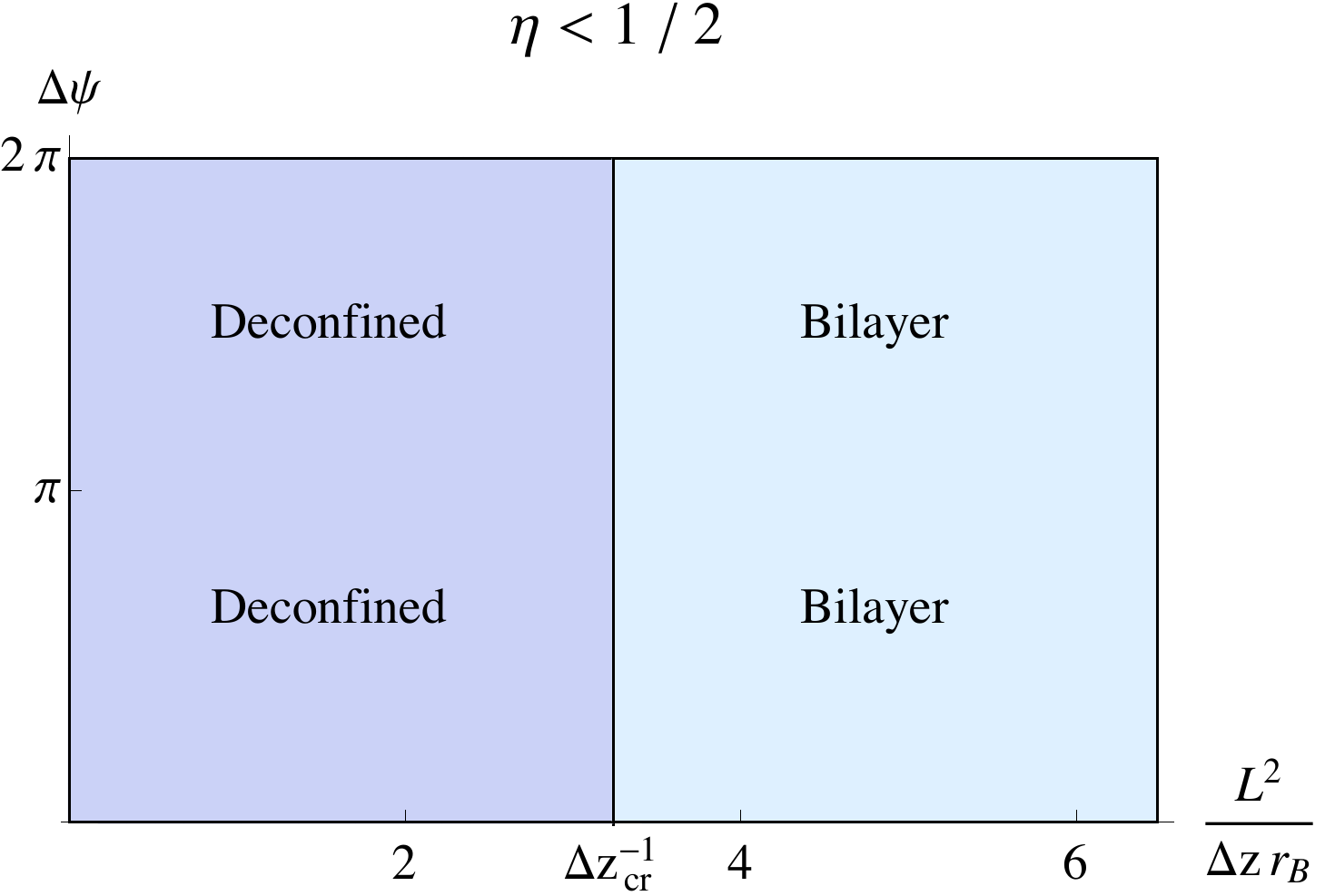}
                              \includegraphics[width=3.4in]{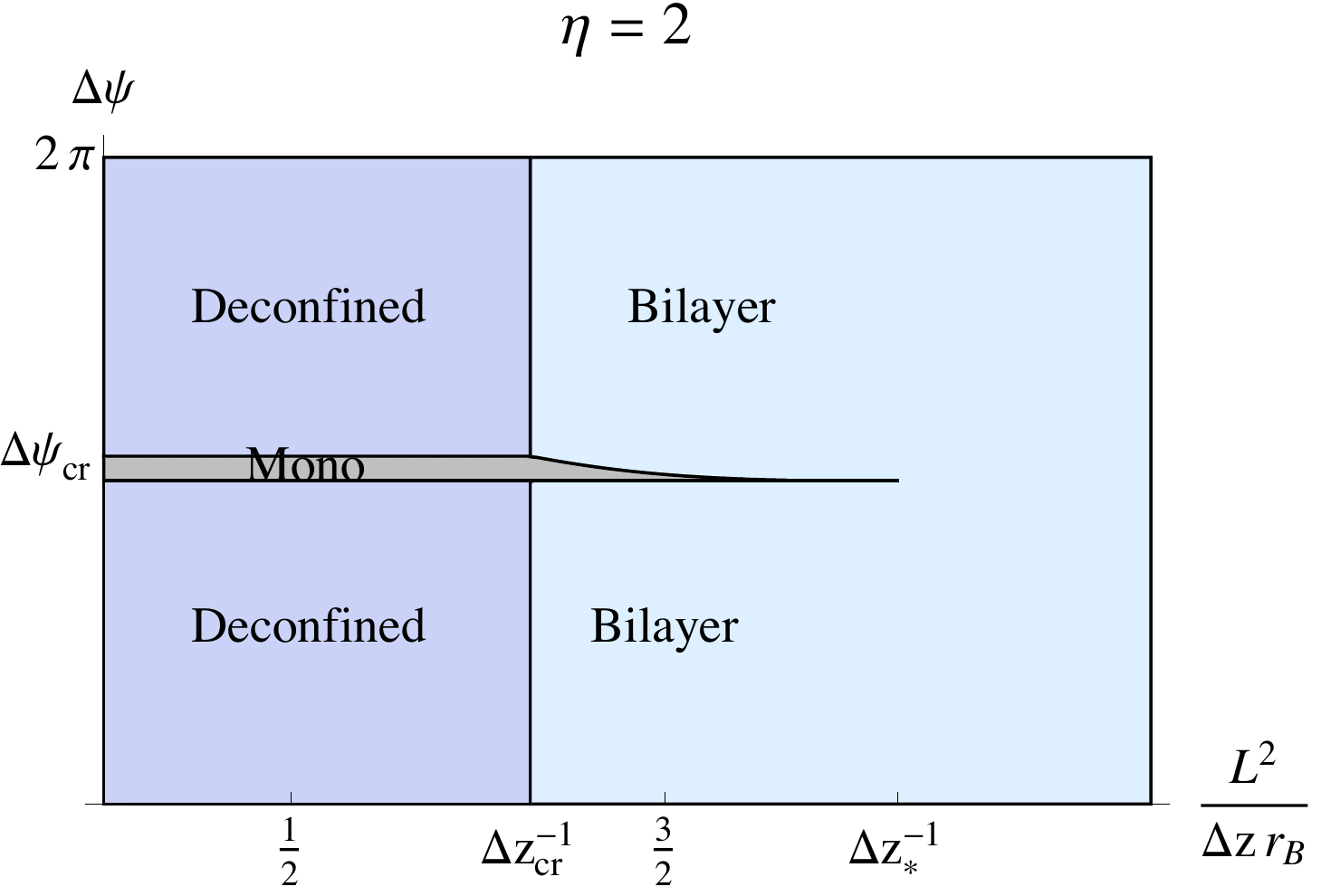}
    
 \vspace{.5cm}   
  \hspace{-0.9cm}  \includegraphics[width=3.4in]{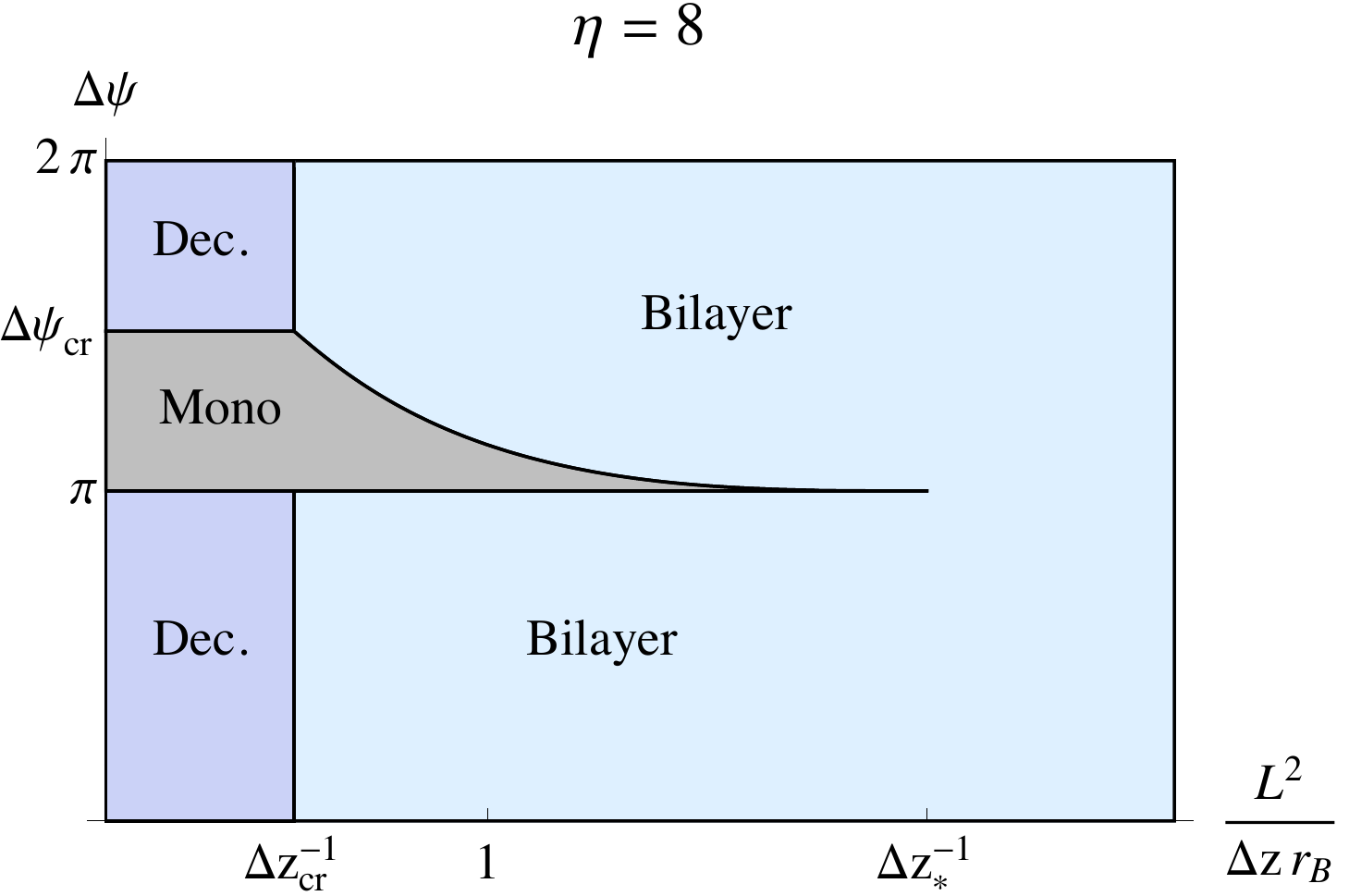}
                              \includegraphics[width=3.4in]{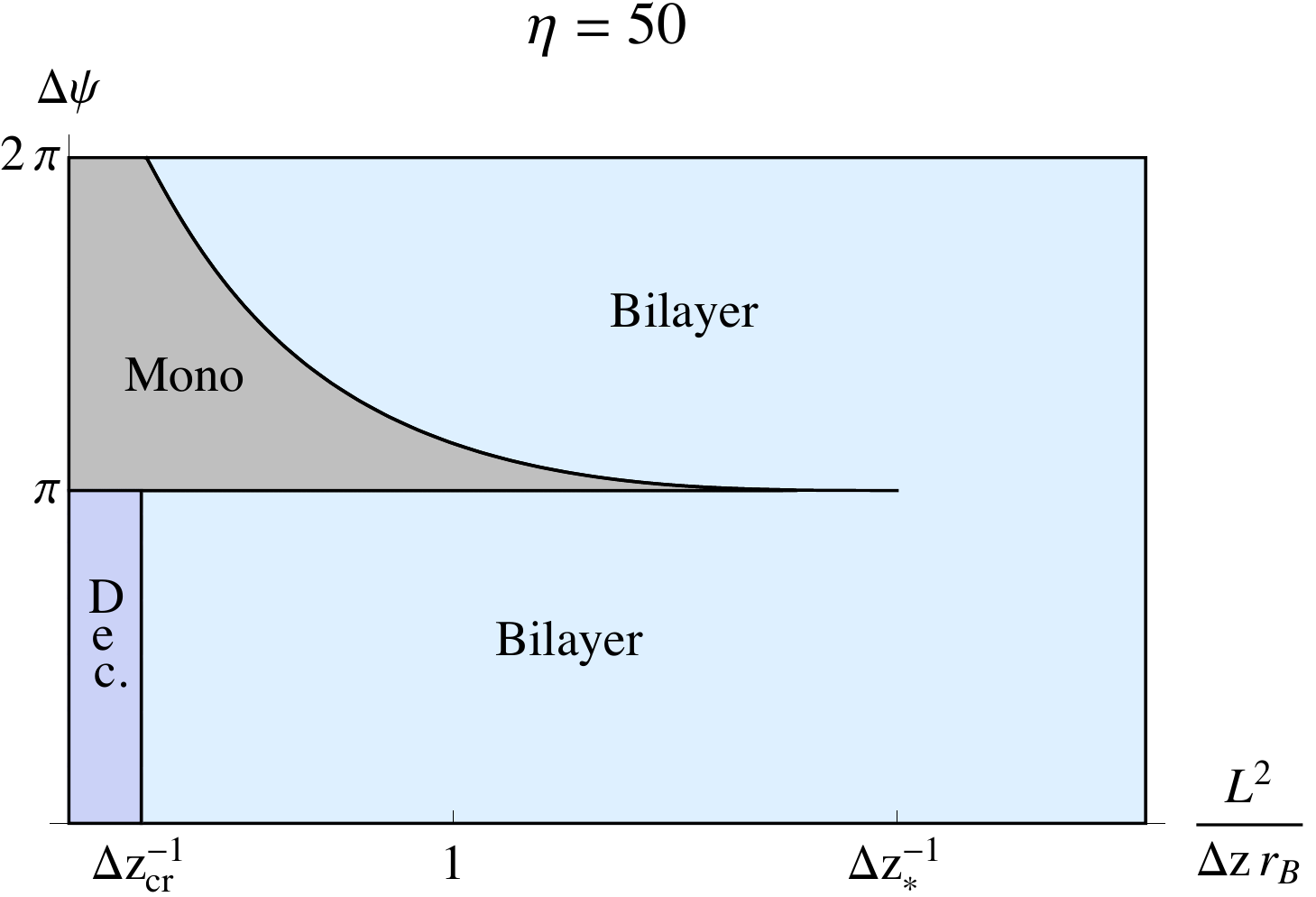}        
   \caption{Phase diagram at finite temperature and finite external magnetic field for various finite values of $\eta$.}
   \label{fig:phaseTBfin2}
\end{figure}

In general the theory has four parameters: temperature ($T_H$), magnetic field ($B$), separation between the layers ($\Delta z$) and the dimensionless parameter $\Delta\psi$. This suggests that the phase diagram can be described by three dimensionless parameters. To facilitate comparison to the zero temperature diagram presented in figure \ref{fig:phaseT0Bfin}, we choose these physical parameters\footnote{These parameters can be chosen as boundary conditions in the dual field theory in the UV.} to be $\Delta\psi$, $L^2/(\Delta z\,r_B)$ and $\eta$, and represent the three dimensional diagram in terms of $\eta=\rm const.$ slices (see figure \ref{fig:phaseTBfin2}).  

According to the analysis carried out in ref.~\cite{Filev:2013vka}, the monolayer phase exists only for sufficiently large ratio of the magnetic field and the temperature squared, namely for $\eta>1/2$.  Therefore, for $\eta<1/2$ the phase diagram in the $\Delta\psi$ -- $L^2/(\Delta z\,r_B)$ plane is determined by the phase transition of the single bilayer phase described in section \ref{SBLTB}. As one can see from figure \ref{fig:phaseTBfin2}, in this case the phase diagram has only deconfined and bilayer phases separated by a critical line of first order phase transitions at $\Delta z_{\rm cr}$. As long as $\eta <1/2$, the evolution of $\Delta z_{\rm cr}$ with $\eta$ is determined by the phase diagram of the single bilayer (see figure \ref{fig:phaseTBsingle}).  

For $\eta>1/2$, the monolayer phase is possible and the phase diagram contains three different phases. As one can glean from figure \ref {fig:phaseTBfin2}, for $\eta=2$ the phase diagram has a strip of monolayer phase, which is stable for $\pi<\Delta\psi\leq\Delta\psi_{\rm cr}$ and $\Delta z > \Delta z_*$. Here, as in the previous section, $\Delta z_*$ is determined by the vanishing of the regularised free energy $F_{U_z}$. Furthermore at $\Delta z = \Delta z_{\rm cr}$ and $\Delta\psi=\Delta\psi_{\rm cr}$, there is a triple point, where all three phases coexist. For $\Delta z > \Delta z_{\rm cr}$ there are critical curves at $\Delta\psi=\Delta \psi_{\rm cr}$ and $\Delta\psi=\pi$, which separate the monolayer and  the deconfined phases. 
For $\Delta z< \Delta z_{\rm cr}$ there are critical curves at $\Delta\psi (\Delta z)$ and $\Delta\psi=\pi$, which merge at $\Delta z=\Delta z_*$ and separate the monolayer and the bilayer phases. From figure \ref{fig:phaseTBfin2}, one can see that as $\eta$ increases the area of the monolayer phase increases as well as the critical parameters $\Delta\psi_{\rm cr}$ and $\Delta z_{\rm cr}$. At the same time the parameter $\Delta z_*$ remains almost unchanged. At sufficiently large $eta$ (at $\eta\approx 44.594$), $\Delta\psi_{\rm cr}=2\pi$ and for even larger $eta$ (see figure \ref {fig:phaseTBfin2} for $\eta=50$) the monolayer phase is the only stable phase for $\Delta\psi>\pi$ and $\Delta z >\Delta z_{\rm cr}$. Finally, in the limiting case $\eta\to\infty$, corresponding to the zero temperature limit, one has $\Delta z_{\rm cr}\to\infty$, the deconfined phase ceases existence and one recovers the phase diagram presented in figure \ref{fig:phaseT0Bfin}.

In conclusion, we see that at finite temperature and external magnetic, the theory has a rich phase structure, characterised by a critical point of three coexisting phases. By dialling the parameters of the theory one can stabilise any of the deconfined, monolayer or bilayer phases.  

\section{Conlusions and outlook}
In the present work, we have explored the phase structure of bilayer and monolayer phases in the Klebanov-Witten model with embedded $D5/\overline{D5}$ flavour probe brane pairs, resulting in a dual $2+1$-dimensional defect field theory of strongly coupled fermions living on domain walls in the $3+1$-dimensional ambient space-time. The main advantage of the Klebanov-Witten background, as opposed to the $AdS_5Ê\times S^5$ background, is the fact that the $D5/\overline{D5}$-brane configurations are stable without the necessity to stabilise them (see \cite{Grignani:2012qz} for a related construction involving $D7/\overline{D7}$-branes in the $AdS_5Ê\times S^5$ background, where it is necessary to apply a pressure at the UV boundary to prevent the branes from annihilating and to keep them at a fixed separation). \vskip0.3cm
We have found a fairly rich phase structure at finite temperature and external magnetic field from the competition of the dissociating effect of the temperature and the 
binding effect of the magnetic field. The major novelty here is that we also find a competition between the two possible U-shaped configurations corresponding to the bilayer (U-shaped embedding in the $z$-direction) and  monolayer (U-shaped embedding in the internal angle $\psi$) phases. The hope is that the results reported here can be used to improve our understanding of spontaneous symmetry breaking in graphene and other $2+1$-dimensional materials. In addition, the bilayer/monolayer phase transitions that we describe could be relevant for real condensed matter systems exhibiting bilayer structures.\vskip0.3cm
Interesting directions for ongoing and future research include the generalisation of the $D5/\overline{D5}$-probe brane embeddings \cite{Ihl:2014} to the Klebanov-Strassler model \cite{Klebanov:2000hb}, its baryonic branch \cite{Butti:2004pk} (or even the non-supersymmetric baryonic branch \cite{Bennett:2011va}), which are appealing for phenomenological reasons, and would potentially allow to study quantum Hall states (cf. e.g, \cite{Kristjansen:2012ny, Kristjansen:2013hma, Bergman:2010gm, Jokela:2011eb}) in this framework, due to the existence of the $C_{(2)}$ RR-form and the corresponding Chern-Simons term in these backgrounds.
It would also be interesting to study other phenomena in this model, like the existence and temperature dependence of zero sound and diffusion modes at finite baryon chemical potential (see e.g. \cite{Davison:2011ek},\cite{DiNunno:2014bxa} for relevant results in other holographic models).

\acknowledgments
The authors want to express their gratitude to Nick Evans and Gordon Semenoff for very useful conversations and correspondence, and for reading the manuscript. M.~I. and D.~Z.~are funded by the FCT fellowships SFRH/BI/52188/2013 and SFRH/BPD/62888/2009, resp.
The Centro de F\'isica do Porto is partially funded by FCT through the projects PTDC/FIS/099293/2008 and CERN/FP/116358/2010.

\end{document}